\documentclass[aps,reprint,pre,showkeys]{revtex4-1}
\usepackage{graphicx}
\usepackage{amsmath}

\newcommand{\DD}{{\mathcal {D}}}
\newcommand{\XX}{{\mathcal X}}
\newcommand{\dRSq}{{\overline{\delta R^2}}}
\newcommand{\muk}{{\mu k}}
\newcommand{\ik}{{i k}}
\newcommand{\mutwo}{{\mu 2}}
\newcommand{\itwo}{{i 2}}
\newcommand{\vv}{{\mathbf{v}}}
\newcommand{\DDelta}{{\mathbf{\Delta}}}

\DeclareMathOperator{\erfc}{erfc}

\begin{document}
\title{Innovation rather than improvement: a solvable high-dimensional model\\ highlights the limitations of scalar fitness}
\author{Mikhail Tikhonov}
    \affiliation{School of Engineering and Applied Sciences; Kavli Institute for Bionano Science and Technology, Harvard University, Cambridge, MA 02138, USA.\\\emph{Present address:} Department of Applied Physics, Stanford University, Stanford, CA 94305, USA}
\author{Remi Monasson}
    \affiliation{Laboratoire de Physique Th\'eorique de l'\'Ecole Normale Sup\'erieure -- UMR 8549, CNRS and PSL Research, Sorbonne Universit\'e UPMC, 24 rue Lhomond, 75005 Paris, France}

\begin{abstract}
Much of our understanding of ecological and evolutionary mechanisms derives from analysis of low-dimensional models: with few interacting species, or few axes defining ``fitness''. It is not always clear to what extent the intuition derived from low-dimensional models applies to the complex, high-dimensional reality. For instance, most naturally occurring microbial communities are strikingly diverse, harboring a large number of coexisting species, each of which contributes to shaping the environment of others. Understanding the eco-evolutionary interplay in these systems is an important challenge, and an exciting new domain for statistical physics. Recent work identified a promising new platform for investigating highly diverse ecosystems, based on the classic resource competition model of MacArthur. Here, we describe how the same analytical framework can be used to study evolutionary questions. Our analysis illustrates how, at high dimension, the intuition promoted by a one-dimensional (scalar) notion of fitness can become misleading. Specifically, while the low-dimensional picture emphasizes organism cost or efficiency, we exhibit a regime where cost becomes irrelevant for survival, and link this observation to generic properties of high-dimensional geometry.
\end{abstract}
\maketitle

The image of evolution proceeding via ``improvement'' is such a convenient metaphor that, although clearly known to be wrong, it still influences our intuition. Mutations make organisms not better, but different; whether such differences are beneficial depends on the context, which itself is subject to change. The issue becomes especially relevant in an ecological context, where the environment of any one organism is defined, among other factors, by the presence and activity of other species, and can therefore change as fast or faster than the timescale of evolutionary processes.

The lack of timescale separation is particularly pronounced in microbial ecology, and the subject of eco-evolutionary interplay is increasingly in the spotlight today in the context of microbiome research~\cite{Fussmann07,Pelletier09,Henson15,Schoener11,Hendry}. The issue is exacerbated by the observation that most naturally occurring microbial communities are highly diverse, harboring a large number of coexisting species~\cite{HMP,EMP}, each of which contributes to shaping the environment of others~\cite{LevinsLewontin,DeMartino,ScottPhillips14,PRL,Advani}. The environment is therefore an intrinsically high-dimensional object, and it is well established that large dimensionality of a problem can lead to qualitatively novel effects~\cite{PWA}.

Developing a theoretical understanding of ecology and evolution in the high-diversity regime is therefore an important challenge. So far, however the barrier of dimensionality proved difficult to cross. Although the improvement metaphor is very clearly understood to be misleading~\cite[and many others]{Guyer,Ayala,Mayr}, most quantitative results in evolutionary theory have been derived in a simplistic one-dimensional picture (the ``fitness landscape'' of classic population genetics), which inevitably reinforces certain expectations. It is not always clear to what extent the intuition derived from low-dimensional models applies to the complex, high-dimensional reality~\cite{LevinsLewontin,DeMartino,PRL,Advani,Levins66,Levin92,Lawrence,FisherMehta,eLife,Bunin,Doebeli,BarbierBunin}.

Recent work identified a promising platform for investigating highly diverse ecosystems using statistical physics of disordered systems~\cite{StatMechOfLearning}, based on the classic resource competition model of MacArthur~\cite{MacArthur,Tilman82,Chesson90}. Resource depletion is the simplest form of feedback of organisms onto their environment, and the high-diversity limit of this model was recently shown to be analytically tractable~\cite{DeMartino} (see also~\cite{PRL,Advani,Barbier}). As the number of resources becomes large, the community was shown to acquire increasing control over the immediate environment experienced by its members~\cite{DeMartino,PRL,Advani}. Clearly, this should have important implications for how evolution would act in such a community~\cite[and references therein]{Lawrence,Doebeli}. Uncovering such implications in this model is the focus of this work. Using the high-diversity MacArthur model as our platform, we identify some important deviations from the low-dimensional intuition. We then argue that our conclusions are in fact more general that this particular model, namely that the breakdown of the improvement metaphor at high diversity stems directly from the properties of high-dimensional geometry.

We begin by briefly defining the model. We largely follow the notations of~\cite{PRL}, but will point out how the model used there for a purely ecological discussion can be applied to study evolutionary questions.

\section{The model.}
Consider a multi-species community in a well-mixed habitat where a single limiting element $\XX$ exists in $N$ forms (``resources'' $i\in\{1\dots N\}$). For example, this could be carbon-limited growth of bacteria in a medium supplied with $N$ sugars. Let $n_\mu$ denote the population size of species $\mu\in\{1\dots S\}$. Briefly, MacArthur's model of resource competition can be described as a feedback loop: The availability $h_i$ of each resource $i$ determines the dynamics of $n_\mu$. The changes in species abundance translate into changes in the total demand for resources, denoted $T_i$. This total demand, in turn, depletes resource availability $h_i$.

Assuming for simplicity that resource dynamics are faster than changes in species abundance, we can assume that resource availability $h_i$ at any moment quickly equilibrates to reflect the instantaneous demand $T_i$ at that same moment: $h_i=h_i(T_i)$. For concreteness, we will posit that all organisms are sharing a fixed total influx of resource $R_i$, so that $h_i(T_i)=R_i/T_i$, with resource supply $R_i=1+\frac{\delta R_i}{\sqrt N}$; this choice of scaling is discussed in Appendix~\ref{app:Model} (see also Ref.~\cite{DeMartino}). In this small-fluctuation regime, the specific functional form of the relation $h_i(T_i)$ is not important as long as its linearization around equilibrium has a decreasing slope~\cite{PRL}, which is a natural condition for stability (higher demand should result in lower supply). It is worth noting that even the large-fluctuation regime of the MacArthur model can be rendered analytically tractable~\cite{Advani}; however, for our purposes the above model is sufficient. The variance of $\delta R_i$ over $i$ is a control parameter describing the heterogeneity of resource supply, and is denoted $\dRSq$.

For a given set of $S$ species, the ecological dynamics we consider reproduce those of Ref.~\cite{PRL}:
\begin{equation}\label{eq:surplus}
\begin{aligned}
\frac{dn_\mu}{dt} &= n_\mu\Delta_\mu(h_i)\\
h_i &= \frac{R_i}{T_i(n_\mu)},
\end{aligned}
\end{equation}
where $\Delta_\mu(h_i)\equiv \sum_i \sigma_{\mu i}\, h_i - \chi_\mu$ (``resource surplus''), and $T_i(n_\mu)\equiv \sum_\mu \sigma_{\mu i}\, n_\mu$ (``total demand'').

In these equations, a species $\mu$ is characterized by its metabolic strategy $\{\sigma_{\mu i}\}$ and its requirement $\chi_\mu$ for the limiting element $\XX$. The population growth rate of species $\mu$ is determined by its resource surplus $\Delta_\mu$. In the expression for $\Delta_\mu$ above, the first term is the total harvest of $\XX$ from all sources, and the second is the requirement an individual must meet to survive; $R_i$ is the total influx of resource $i$. At equilibrium, each species is either absent ($n_\mu=0$), or its resource intake and expenditure are balanced ($\Delta_\mu=0$). A stable (\emph{non-invadeable}) equilibrium is characterized by an extra condition that all the absent species, if introduced, would be driven back to extinction: if $n_\mu=0$, then $\Delta_\mu<0$. The dynamics defined above always has a stable equilibrium, uniquely defined~\cite{PRL,eLife,MacArthur} by the set of competing species.

We have defined the \emph{ecological} dynamics for a given set of species. To specify an \emph{evolutionary} process, we now need to describe how this set is constructed and evolves. In our model, we posit that evolution ``discovers'' random new species one by one, and these are added to the pool of competitors. Each new species has a random strategy vector $\vec\sigma^*$, which we take to be binary for simplicity (each entry $\sigma^*_{i}$ is 1 with probability $p$ and 0 otherwise), and a random cost $\chi^*=\sum_i\sigma^*_i+\epsilon x^*$ with a normally distributed $x^*$. This cost model corresponds to the assumption of approximate neutrality~\cite{PRL,eLife,Neutral}, and is discussed in Appendix~\ref{app:Model}. We assume that new species are generated sufficiently slowly that the ecosystem has time to equilibrate before a new species is introduced. The ``evolutionary sequence'' we consider is the sequence of non-invadeable ecological equilibria resulting from the competition of all species discovered up to that moment. In other words, we consider the increasing number of species in the pool $S$ as a proxy for evolutionary time. It is convenient to normalize $S$ by $N$, defining a parameter $\alpha=S/N$. We will say that a species is ``alive (dead) at time $\alpha$'' if it is present (absent) at the respective equilibrium.

To summarize, the control parameters of the model include $N$ (the number of resources), $\alpha$ (the size of the species pool, serving as a proxy for evolutionary time), $\epsilon$ (the width of the cost distribution), $p$ (the sparsity of the metabolic strategy vectors), and the heterogeneity of resource supply $\dRSq$. All simulations below use $\epsilon = 10^{-4}$, $p = 0.5$, and are performed as described in Ref.~\cite{PRL}. Parameters $N$ and $\dRSq$ are specified in the respective legends, and $\alpha$ is taken as the variable against which all results are plotted. A Matlab script reproducing all figures is provided as Supplementary file 1.

It is worth stressing the simplifying assumptions made above. First, we take each new species to be fully random, rather than a small modification of an existing one. Second, the particular evolutionary process we consider proceeds through a sequence of equilibria. These assumptions make the model analytically tractable while preserving the main feature of interest, namely the eco-evolutionary feedback in a high-dimensional environment, and so provide a reasonable starting point for investigation. Finally, our cost model assumes approximate neutrality, where the effects we seek are likely strongest. Ideally, such effective neutrality should itself be exhibited as an outcome of an evolutionary process; here, we treat it as an empirically-motivated assumption (see Appendix~\ref{app:Model}).

\begin{figure*}
  \includegraphics[width=0.8\textwidth]{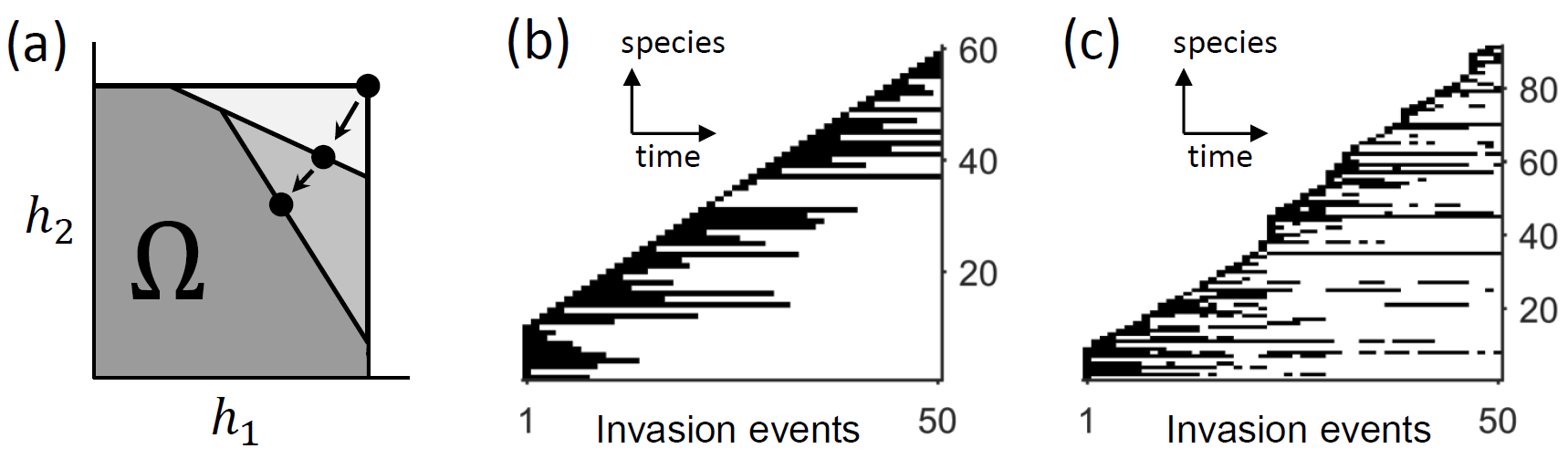}
\caption{\textbf{(a)} The geometric interpretation of MacArthur's model. In the space of resource availability, a given set of species defines a convex region $\Omega$ where no species can survive. The dynamical equilibrium is always located at the boundary of $\Omega$, and successful invaders slice off regions of $\Omega$. In this cartoon, $N=2$ and a community of two specialists $(1,0)$ and $(0,1)$ is invaded first by one generalist $(0.7, 0.3)$, and then by another $(0.4, 0.6)$, with progressively lower costs; arrows connect successive equilibria. (For the purposes of illustration, here we allow arbitrary metabolic strategies, without enforcing the constraint $\sigma_{\mu i}\in\{0,1\}$, as in the text.) This geometric intuition suggests that evolution proceeds towards stronger depletion of resources, and consists of organisms with progressively smaller costs, which can then be interpreted as higher efficiency. One might expect such evolutionary process to be qualitatively similar to the simple ``best $N$ species'' model, shown in panel \textbf{(b)}. The ``best $N$ species'' model ignores any metabolic considerations; instead, each newly generated species is simply assigned a ``fitness'' value, and the community consists of the $N$ species with highest fitness. The panel shows species presence (in black; each species is a row) for 50 invasion events in one random realization of such a model with $N=10$. Each species is present for a period of time, until outcompeted by a more efficient species. \textbf{(c)} The actual simulations of MacArthur model ($N=15$, $\dRSq=1.5$) show a very different pattern of behavior. Most strikingly, the outcompeted species routinely return to again be present at equilibrium.}
\label{fig:1}
\end{figure*}

Our model has a convenient geometric interpretation, first introduced by Tilman~\cite{Tilman82} and illustrated in Fig.~\ref{fig:1}a. Following Ref.~\cite{PRL}, we can think of the metabolic strategies $\{\sigma_{\mu i}\}$ as $S=\alpha N$ vectors in the $N$-dimensional space of resource availability. Below each hyperplane $\vec h\cdot\vec \sigma_\mu=\chi_\mu$ lies the half-space where available resources are insufficient to support species $\mu$. The intersection of such regions over all competing strategies $\{\vec\sigma_\mu, \chi_\mu\}$ defines the ``unsustainable region'', which we denote $\Omega$:
\begin{equation}\label{eq:Omega}
\Omega = \bigcap_{\mu=1}^{\alpha N}\; \{\vec h \;|\; \vec h \cdot \vec\sigma_\mu<\chi_\mu\}
\end{equation}
When resource availability $\vec h$ is inside $\Omega$, no species can harvest enough resources to sustain its population. Outside $\Omega$, at least one species can increase its abundance. Therefore, the equilibrium state can only be located at the boundary of $\Omega$, and can in fact be found by solving an optimization problem over this region~\cite{PRL}.

This geometric picture proved highly influential, and was used to study competition for $N=1$ and $N=2$ resources in great detail~\cite[and others]{Tilman82}. It also provides a clear intuition for the evolutionary sequence in our model: as new species are generated, some will have a cost low enough to allow them to invade the community, slicing another piece off the unsustainable region (Fig.~\ref{fig:1}a). As a result, resource depletion becomes progressively stronger, and community composition shifts towards progressively more efficient species, where efficiency is measured as the species' cost per pathway. In fact, the dynamics~\eqref{eq:surplus} has a Lyapunov function, and each successive equilibrium corresponds to a lower value of $\sum_i R_i \log h_i$, precluding any rock-paper-scissors scenarios, and reinforcing the expectation of a ``linear'' progression~\cite{eLife}.

This intuitive picture emphasizes the role played by the individual species' efficiency, and suggests that the evolutionary sequence should be qualitatively similar to the ``best $N$ species'' model illustrated in Fig.~\ref{fig:1}b. This is a much simpler model where every species is described by a single value one calls ``fitness'', and at each point in time, the community consists of the best $N$ species ``discovered'' to date. In other words, this model is a natural generalization of ``survival of the fittest'' to the case where $N$ resources allow coexistence of up to $N$ species. In the ``best $N$'' model, each species enjoys a period of existence, until it is outcompeted by someone better.

However, the expectations set up by a low-dimensional picture often prove incorrect at high dimension. Fig.~\ref{fig:1}c shows an evolutionary sequence for one simulation of the MacArthur model at $N=15$. The striking qualitative difference between Fig.~\ref{fig:1}b and Fig.~\ref{fig:1}c is that the supposedly ``outcompeted'' species keep returning to again become part of the equilibrium state.

In our model, species are never removed from the pool of competitors. Although a particular species may be driven to zero abundance at a certain equilibrium state, it never goes completely extinct: biologically speaking, we assume that each previously discovered species is preserved in some other spatial patch, or in a dormant form (a spore). Thus, the model is set up to allow a previously outcompeted species to keep trying its luck again at subsequent times. Nevertheless, such ``returns from the dead'' appear surprising.

A clear qualitative phenomenon that contradicts the naive intuition is a good window into how evolution might be acting differently at high diversity. We therefore start by characterizing this effect within our model: for a species that was alive at time $\alpha_1$ but dead at $\alpha_2$, what is the probability of its return at time $\alpha_3$? (Once again, in our model, the number of species in the competitor pool serves as a proxy for evolutionary time.)

\section{The replica-theoretic calculation}

MacArthur model is special in that it is a global optimization problem~\cite{MacArthur}, and this makes it analytically tractable using methods of statistical physics of disordered systems~\cite{DeMartino,PRL,Advani,StatMechOfLearning}. At equilibrium, we expect $h_i\approx 1$ (see Appendix~\ref{app:Model}), and we therefore introduce:
\[
g_i\equiv 1-h_i.
\]
The order parameters are:
\[
\begin{aligned}
m&=\sum_i g_i\\
q&=\sum_i g^2_i
\end{aligned}
\]
Both $m$ and $q$ can be explicitly computed as functions of the control parameters of the model.

\begin{figure*}
  \includegraphics[width=0.8\textwidth]{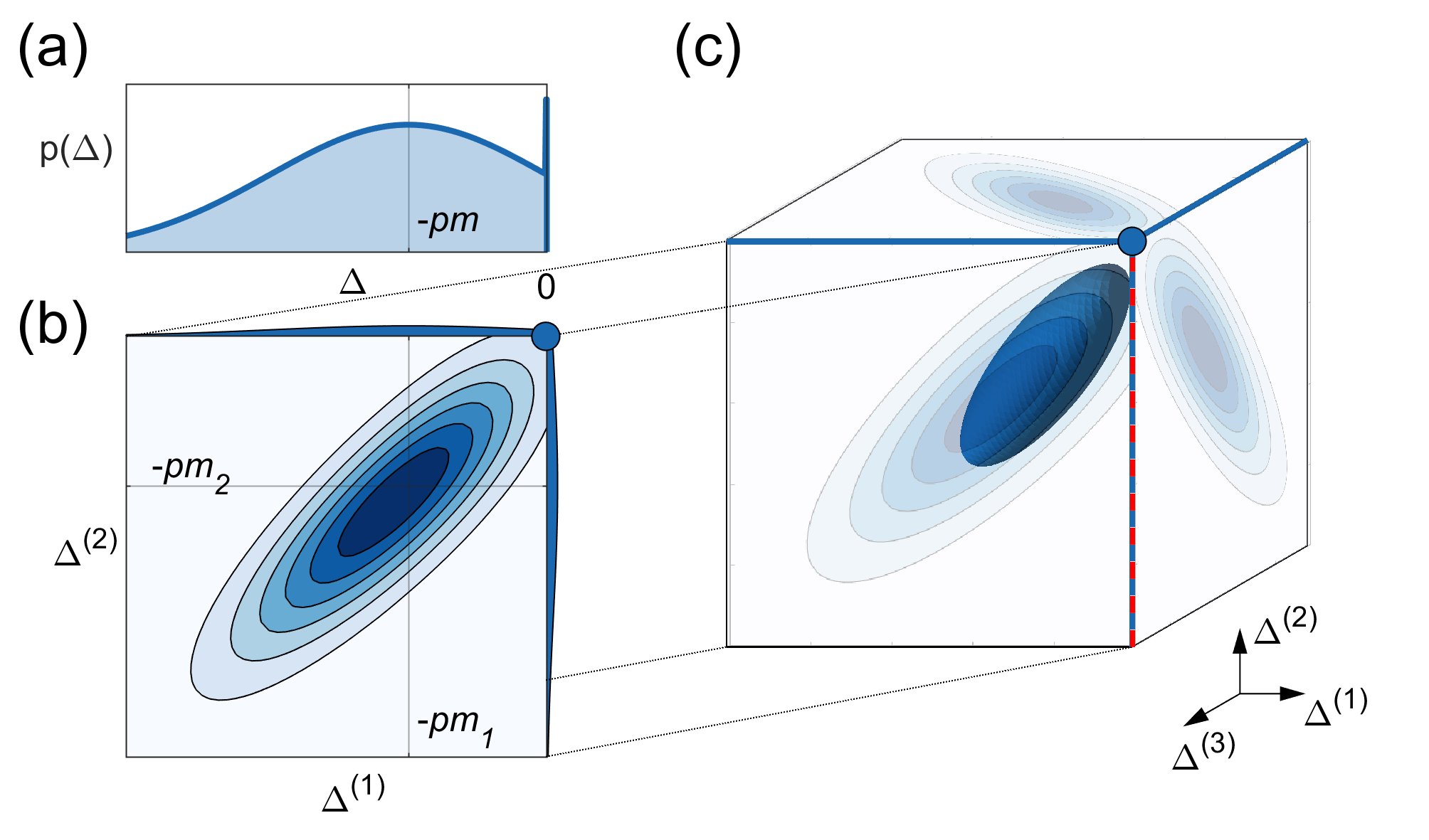}
\caption{\textbf{The structure of the joint distribution $p_{\alpha_1\dots\alpha_k}(\Delta^{(1)}\dots\Delta^{(k)})$.}
\textbf{(a)} For $k=1$, the probability distribution $p_\alpha(\Delta)$ is a Gaussian whose weight over the positive semi-axis is collected into a delta-peak corresponding to the survivors~\cite{PRL}. \textbf{(b)} For $k=2$, the 2-point distribution is a 2-dimensional Gaussian restricted to the negative quadrant, with excess probability accumulating at the quadrant boundaries. The projection of this distribution on any of the two axes takes the shape shown in (a), as it must. Knowing the marginals leaves one free parameter to be determined through a replica-theoretic calculation, corresponding to the correlation of $\Delta^{(1)}$ and $\Delta^{(2)}$. \textbf{(c)} For $k=3$, the shape of the distribution is again a multi-dimensional Gaussian restricted to negative $\Delta^{(1),(2),(3)}$. Importantly, for all $k\ge3$ such a distribution is fully determined by its known 2-dimensional projections (gray lines).}
\label{fig:2}
\end{figure*}

The surviving species are those whose resource surplus $\Delta$ is exactly zero. For any one $\alpha$, the probability distribution $p_\alpha(\Delta)$ was also computed in Ref.~\cite{PRL} and is shown in Fig.~\ref{fig:2}a. Its Gaussian part is centered at $-pm$ with variance $p(1-p)q+\epsilon^2$, and the delta-peak corresponds to the survivors.

To understand evolutionary sequences in our model, we need to know what happens if new species are added to the system. We know the new number of survivors, but how many of these were already present and stayed on, and how many are newcomers? This information is encoded in the joint probability $p_{\alpha_1\dots\alpha_k}(\Delta^{(1)}\dots\Delta^{(k)})$, computed for a typical evolutionary sequence. In particular, the return probability of a previously outcompeted species (alive at $\alpha_1$, dead at $\alpha_2$, alive again at $\alpha_3$) is encoded is the three-point joint probability $p_{\alpha_1\alpha_2\alpha_3}$.

Before tackling the 3-point problem, let us begin by computing the 2-point joint probability $p_{\alpha\alpha'}(\Delta,\Delta')$. To do so, consider two copies of a system, where the first has $\alpha N$ species, while the second has those same species plus $(\alpha'-\alpha)N$ extra ones. The calculation is similar to that of Ref.~\cite{PRL}, except in addition to $m$, $q$ and $m'$, $q'$ characterizing each of the two copies (and satisfying the same equations as in Ref.~\cite{PRL} for respectively $\alpha$ and $\alpha'$), there is now an additional order parameter describing the coupling:
\[
r=\sum_i g_i g'_i
\]
The replica calculation is relatively straightforward and is detailed in Appendix~\ref{app:R}.
We find that the 2-point distribution $p_{\alpha\alpha'}$ is a double Gaussian restricted to the negative quadrant (Fig.~\ref{fig:2}b), with excess probability accumulating at the quadrant boundaries (corresponding to species alive only at $\alpha$ or only at $\alpha'$) and at the origin (corresponding to species that are alive at both $\alpha$ and $\alpha'$).
Just as the variance of $p_\alpha(\Delta)$ was given by $p(1-p)q+\epsilon^2$, we find that the correlation matrix of the Gaussian in $p_{\alpha\alpha'}(\Delta,\Delta')$ is given by:
\[
C=
p(1-p)\begin{pmatrix}
  q & r \\
  r & q'
\end{pmatrix}+\epsilon^2\begin{pmatrix}
  1 & 1 \\
  1 & 1
\end{pmatrix}
\]

The truncated double-Gaussian form of $p_{\alpha\alpha'}(\Delta,\Delta')$ was expected, since the two marginals of this distribution $p_\alpha(\Delta)$ and $p_{\alpha'}(\Delta')$ must take the form shown in Fig.~\ref{fig:2}a. The key parameter to determine from the replica calculation is $r$. For a homogeneous resource supply ($\dRSq=0$, i.e. all $R_i$ are identical), the equation we find takes an especially simple form:
\begin{equation}\label{eq:r}
r = \frac{\alpha}{p(1-p)}\int_{0}^\infty\DD(\omega,\omega')\,\omega\omega'
\end{equation}
Here $\omega$, $\omega'$ are auxiliary noise variables, and $\DD(\omega,\omega')$ is a 2-dimensional Gaussian measure with the correlation matrix $C$ above:
\[
\DD(\omega,\omega') =
\frac{d\omega\,d\omega'}{2\pi\sqrt{\det C}}\exp\left\{-\frac12
\left(\begin{smallmatrix}\omega+p m\\\omega'+p m'\end{smallmatrix}\right)^t
[C(r)]^{-1}
\left(\begin{smallmatrix}\omega+p m\\\omega'+p m'\end{smallmatrix}\right)\right\}
\]
We use the notation $C(r)$ to stress that the unknown $r$ enters on both sides of the equation~\eqref{eq:r}. For a given set of parameters, this equation can be solved numerically. The general equation for a heterogeneous resource supply ($\dRSq\neq 0$) and its derivation is provided in Appendix~\ref{app:R}.

We now return to the problem of computing the complete multi-point joint distribution.
Thanks to the mean-field nature of our model, and similarly to the 2-dimensional case, the 3-point joint distribution $p_{\alpha_1\alpha_2\alpha_3}$ can only take the form of a 3-dimensional Gaussian restricted to the negative octant $\Delta^{(1),(2),(3)}<0$ (Fig.~\ref{fig:2}c). 
Now, however, this distribution is entirely determined by its projections, with no free parameters. Our 2-point calculation above is therefore sufficient to write the expression for the 3-point (or indeed any $n$-point joint distribution). Specifically, inside the octant the probability distribution is a 3-dimensional Gaussian with the correlation matrix:
\[
C^{(3)}=
p(1-p)\begin{pmatrix}
  q_1 & r_{12} & r_{13}\\
  r_{12} & q_2 & r_{23}\\
  r_{13} & r_{23} &q_3
\end{pmatrix}+\epsilon^2\begin{pmatrix}
  1 & 1 & 1\\
  1 & 1 & 1\\
  1 & 1 & 1
\end{pmatrix},
\]
and excess probability accumulates at the octant boundary. In this expression, the quantities $q_{1,2,3}$ are known from the single-copy calculation~\cite{PRL}, and $r_{ij}$ are found as solutions to the appropriate version of the two-copy equation~\eqref{eq:r}.

Knowing the joint distribution $p_{\alpha_1\alpha_2\alpha_3}(\Delta_1,\Delta_2,\Delta_3)$, the ``return probability'' is easily computed as the conditional probability:
\begin{multline*}
p_{\text{return}}(\alpha_3\:|\:\text{alive at $\alpha_1$, dead at $\alpha_2$})\\=
\frac{p_{\alpha_1\alpha_2\alpha_3}(\Delta_1=0,\Delta_2<0,\Delta_3=0)}{p_{\alpha_1\alpha_2}(\Delta_1=0,\Delta_2<0)}.
\end{multline*}
For $\alpha_1<\alpha_2<\alpha_3$, this is the probability that a previously outcompeted species (alive at $\alpha_1$, dead at $\alpha_2$) will again be present at equilibrium at a later time $\alpha_3$.

\section{The return probability as a window into evolutionary process}
\begin{figure*}[t!]
  \includegraphics[width=\textwidth]{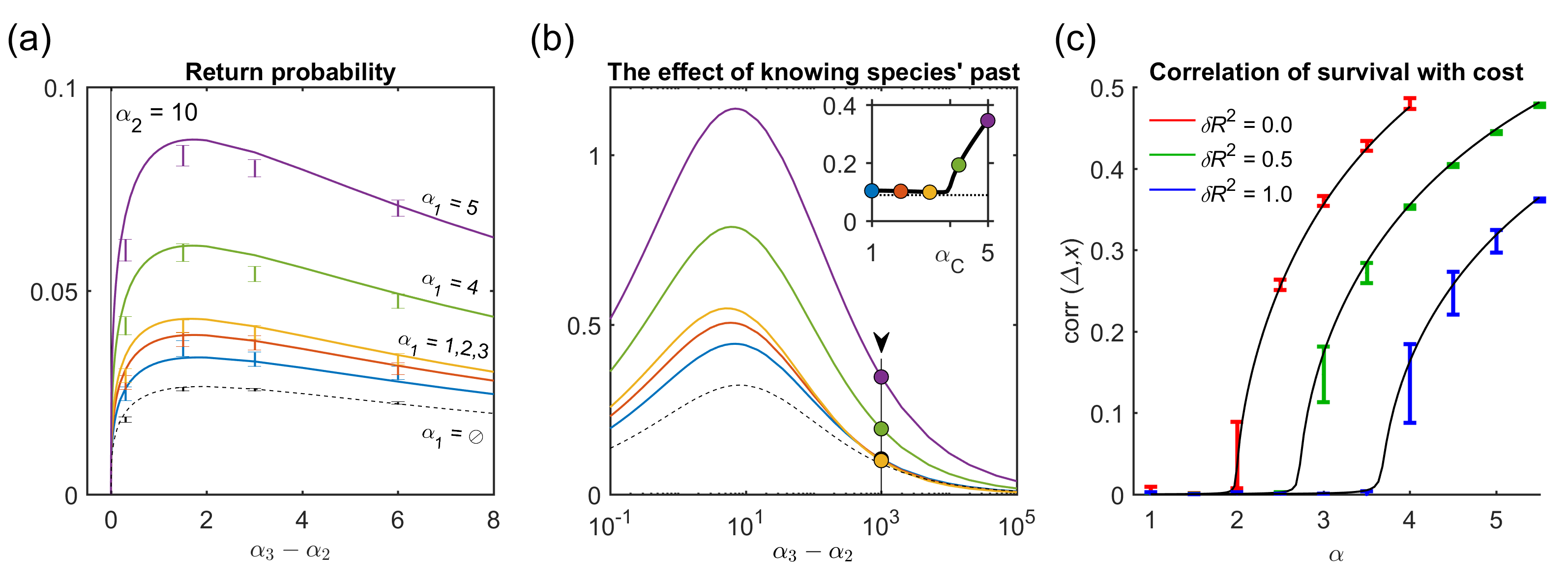}
\caption{\textbf{(a)} The return probability $p_\text{return}(\alpha_3\:|\:\text{alive at $\alpha_1$, dead at $\alpha_2$})$ as a function of $\alpha_3-\alpha_2$, for $\alpha_2=10$ and several values of $\alpha_1$. Theoretical curves (solid lines) agree with simulations ($N=200$; $\dRSq=1$; error bars in both (a) and (c) show standard error of the mean over 100 replicates). Also shown is the 2-point conditional probability of a species absent at $\alpha_2$ to appear at $\alpha_3$ (dotted line labeled $\alpha_1=\oslash$).
\textbf{(b)} Same as (a), normalized by $1/\alpha_3$ (the asymptotic probability of a randomly drawn species to be present at $\alpha_3$) to show long-term behavior. Knowing that a species was absent at $\alpha_2$ makes it less likely to be present at $\alpha_3$ than a randomly drawn species (dotted line, always below 1). Additional information of being alive at $\alpha_1$ enhances the chances [colored lines as in (a)]. For a large $\alpha_3$ (transect) the enhancement remains substantial only if $\alpha_1$ exceeds a critical value $\alpha_C\approx 3.7$ (inset; black line is the theoretical prediction). This critical value corresponds to the V/S phase transition in the model~\cite{PRL}, and signals a change in how selection pressure acts in the two phases.
\textbf{(c)} The correlation of species cost $x$ with resource surplus $\Delta$ (and therefore survival) as a function of $\alpha$. Shown are three curves for different heterogeneity of resource supply $\dRSq$, a parameter that shifts the location of the phase transition (the critical alpha $\alpha_C=2.0$, $2.7$ and $3.7$, respectively). For $\alpha<\alpha_C$, cost and resource surplus are uncorrelated. In this phase cost is irrelevant for survival, in stark contrast to the intuition of Fig.~\ref{fig:1}.}
\label{fig:3}
\end{figure*}

Fig.~\ref{fig:3}a shows $p_{\text{return}}(\alpha_3\:|\:\alpha_1,\alpha_2)$ as a function of $\alpha_3-\alpha_2$ for several values of $\alpha_1$, with $\alpha_2$ fixed at $10$ for all curves. For comparison, the panel also shows the 2-point conditional probability $p(\text{alive at $\alpha_3$}\:|\:\text{dead at $\alpha_2$})$; this corresponds to not specifying any information at $\alpha_1$, and so this curve is labeled ``$\alpha_1=\oslash$''. The analytical results are in good agreement with simulations (Fig.~\ref{fig:3}a). The observed deviations are a small-$N$ effect of simulations. Although we used a value of $N$ as large as computationally feasible ($N=200$), estimating many-point correlation functions from simulations is especially demanding in sample size; but note the excellent match with the $\alpha_1=\oslash$ curve, which is a two-point function.

This agreement allows us to use our analytical results to investigate the large-argument behavior of $p_{\text{return}}(\alpha_3\:|\:\alpha_1,\alpha_2)$. Information about a species' past modifies the expected probability of its survival: knowing it was present or absent a particular time in the past makes our chances to observe it at a future time, respectively, higher or lower. Any consistent ``selection pressure'' (when species favored in the past continue to be favored in the future) should create persistent memory effects. Thus, the long-term behavior of $p_{\text{return}}$ provides important insight into how evolution proceeds in this model.

When $\alpha_3$ is large, $p_\text{return}$ goes to zero and it is convenient to multiply this probability by $\alpha_3$. Indeed, the number of survivors at equilibrium is of order $N$ (in fact, asymptotes to exactly $N$; see~\cite{PRL}), while the number of species in the pool is $\alpha_3N$. The probability of a \emph{randomly drawn} species to be present at equilibrium is therefore $1/\alpha_3$. At large $\alpha_3$, this is a natural baseline against which the return probability should be evaluated. The ratio of $p_{\text{return}}$ to $1/\alpha_3$ quantifies the ``effect of knowing a species' past'', and is shown in Fig.~\ref{fig:3}b.

Consider the dependence of these curves on parameter $\alpha_1$. Qualitatively, the behavior is easy to understand. For fixed $\{\alpha_2,\alpha_3\}$, the return probability increases with $\alpha_1$, and the $\alpha_1=\oslash$ curve is the lowest of all (Fig.~\ref{fig:3}a,b). This is the expected behavior: knowing that a species was previously alive increases its chances of being found alive again, and the effect is stronger if the information is recent.

Quantitatively, the behavior is much more interesting. Keeping $\alpha_2=10$, let us fix $\alpha_3$ to some large value, for instance $\alpha_3=10^3$, and consider $p_\text{return}$ as a function of $\alpha_1$ (the transect in Fig.~\ref{fig:3}b). We find that the return probability exhibits a marked transition in behavior: knowing that a species was present at $\alpha_1$ can greatly increase its likelihood of survival at $\alpha_3$ --- but only if $\alpha_1$ exceeds a critical value $\alpha_1>\alpha_C$ (Fig.~\ref{fig:3}b, inset). This behavior stems from the V/S phase transition described previously in this model, and indicates that the ``selection pressure'' is very different in the two phases. Indeed, one can show that for $\alpha$ below $\alpha_C$, a species' cost has no bearing whatsoever on its survival, in stark contrast with the intuition of Fig.~\ref{fig:1}a. This surprising behavior is demonstrated in Fig.~\ref{fig:3}c which shows the correlation coefficient between a species' cost $x_\mu$ and its resource surplus at equilibrium $\Delta_\mu$. The theoretical curves (derived in Appendix~\ref{app:Corr}) predict zero correlation for $\alpha<\alpha_C$, again in excellent agreement with simulations. Knowing that a species was alive at $\alpha_1<\alpha_C$ tells us nothing about its cost, and so does not modify its likelihood of survival at a large $\alpha_3$, explaining the behavior observed in Fig.~\ref{fig:3}b (inset).

As we can see, the phase transition exhibited by our model endows it with a rich behavior. Although highly interesting, this transition is specific to this particular model and critically relies on the assumption of approximate neutrality, and we will return to its detailed exploration elsewhere. Here, we will focus instead on the general lesson, namely the failure of the intuition suggested by the low-dimensional picture of Fig.~\ref{fig:1}. In the final section, we will trace the origin of this failure to a generic property of high-dimensional geometry.

\section{Improvement vs. innovation}
\begin{figure*}[t!]
  \includegraphics[width=0.8\textwidth]{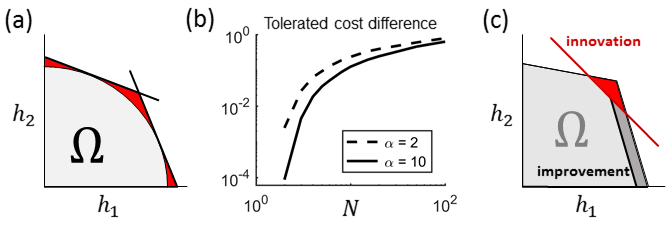}
\caption{\textbf{(a)} The unsustainable region $\Omega$ defined by species with strategies drawn from a unit sphere, all of cost $1$. For $N=2$, most of the region $\Omega$ lies within the unit circle, and any successful invader must have a cost very close to 1. However, as dimensionality increases, the relative volume of the shaded ``corners'' lying outside of the circle grows exponentially, and higher cost differences can be tolerated.
\textbf{(b)} The tolerated cost difference $\delta \chi \equiv \chi-1$ for a random new strategy to be ``viable'' (see text), i.e. to intersect the unsustainable region $\Omega$ defined by $\alpha N$ random strategies of cost 1 as in (a). For each $N$ and $\alpha$, 1000 trials were simulated (sampling 100 random invaders, in 10 independent replicates); shown is the $\delta\chi$ sufficient to ensure viability in 90\% of the cases. Error bars not shown to reduce clutter. At $N=2$ the tolerated $\delta\chi$ is very small, but quickly becomes of order 1 as the dimension increases.
\textbf{(c)} The two paths to invade a community.}
\label{fig:4}
\end{figure*}

What exactly was wrong with the intuition of Fig.~\ref{fig:1}a? The low-dimensional picture places significant emphasis on cost. However, a simple argument illustrates why this emphasis is misleading at high dimension.

Consider a pool of $\alpha N$ species, whose strategy vectors $\vec\sigma$ are drawn uniformly from the $(N-1)$-dimensional sphere, all with cost exactly 1. Together, they define a certain unsustainable region $\Omega$. For any other species, let us call it \emph{viable} if there exists a point in $\Omega$ that can support its growth. Viability of a species means that there exist circumstances (an appropriately chosen vector of resource supply) under which it would be able to invade. For a given, externally fixed supply $\vec R$, viability is required, but not sufficient for invasion.

Drawing a random new strategy $\vec\sigma^*$ uniformly from the same sphere, what cost $\chi$ should we give it to ensure, say, a 90\% probability that this species is viable? Specifically, how high can we go above $\chi=1$?

For $N=2$, the unsustainable region lies almost entirely within the unit circle (Fig.~\ref{fig:4}a), and the maximum tolerated $\delta\chi\equiv\chi-1$ is exceedingly small (Fig.~\ref{fig:4}b; numerical results). However, as dimensionality increases, the relative volume of the shaded regions in Fig.~\ref{fig:4}a explodes; this property is often used to illustrate the counter-intuitive nature of high-dimensional geometry, e.g. by comparing the volume of a cube and its inscribed sphere. As a result, the cost of the strategy becomes essentially irrelevant for viability. As an example, let us take  $\alpha=10$. At $N=2$ ensuring 90\% viability requires a cost of at most $1.00008$, so cost is extremely important: any successful invader is required to be as efficient as the existing community members. But at $N=20$, the threshold becomes 1.25, and so any new species with a cost close to 1 is viable. In this regime, the cost has virtually no role in determining invasion success: whether a species can invade the community depends exclusively on its metabolic strategy.

To put this in intuitive terms, there are two paths to invade a given community (Fig.~\ref{fig:4}c). One is to do what someone is already doing, but better. This corresponds to slicing off a region of $\Omega$ by a cut parallel to an existing plane, and we call this path ``improvement''. The alternative is to slice off one of the many corners of $\Omega$ along a new plane, introducing a strategy distinct from any existing ones (and quite possibly more costly): a path that statistical-mechanics models of economies call ``innovation''~\cite{DeMartino,DeMartino04,Bardoscia}. In dimension 1 (the classic one-dimensional fitness), the innovation path (not reducible to improvement) does not exist, but in high dimension it can easily become the dominant mode of invasion, as we have seen in our model.\cite{Note1}

\section{Discussion}

In this work, we have used an exactly solvable model with eco-evolutionary feedback~\cite{DeMartino} to demonstrate how the intuition derived from a low-dimensional picture fails at high dimension (in our model, the dimension is the number $N$ of resources for which organisms are competing). To the extent that metaphors are useful, we propose that the high-dimensional evolutionary process considered here can be better described through the metaphor of innovation, rather than improvement. Borrowed from the business literature, the term ``innovation'' also seems to carry the connotation of enhancement, but it is often stressed that innovation is rarely focused on cost reduction, but rather on finding new markets (niches), and is often costly, but necessary for survival of an enterprise~\cite{Drucker}. Importantly, innovation is also distinct from invention: while some evolutionary ``discoveries'' result in a qualitative change of lifestyle or physiology (e.g., the evolution of photosynthesis or flight), most need not have that character. As far as metaphors go, the analogy seems appropriate.

In real life, the reasons for the improvement metaphor to be violated are plentiful, and frequently cited. For instance, coevolution of predator and prey can proceed in circles, prey continuously changing its strategy to avoid the predator, and the predator adapting. Other mechanisms include non-transitivity (rock-paper-scissors scenarios; see e.g. Ref.~\cite{Lutz} in this issue), or non-adaptive evolutionary mechanisms such as inherent stochasticity, or hitchhiking mutations. For all these reasons, as is often discussed, it would certainly be naive to expect real ecological or evolutionary dynamics to ever be a simple gradient ascent. It is therefore important to stress once again that all the phenomena described in this work were studied in a model whose global dynamics \emph{does} constitute a gradient ascent, making the initial observation of Fig.~\ref{fig:1} all the more surprising. Our analysis highlights that in high dimension, the intuition promoted by a one-dimensional notion of fitness is likely misleading even if none of the additional mechanisms are at play.

\begin{acknowledgements}
We thank Michael P. Brenner, Andreas Engel, Daniel S. Fisher, Carl P. Goodrich, Alpha Lee, David Zwicker, Harvard Center of Mathematical Sciences and Applications, IESC Cargese and the Simons Foundation. MT was supported in part by National Science Foundation grant DMS-1411694.

\end{acknowledgements}




\newpage
\onecolumngrid
\appendix

 \renewcommand{\thefigure}{A\arabic{figure}}
 \setcounter{figure}{0}

\section{Assumptions of the model}\label{app:Model}
\subsection{Resource fluctuations}
The model considered in this work assumed that the external supply of resources takes the following form:
\begin{equation}\label{eq:scalingR}
R_i=1+\frac{\delta R_i}{\sqrt N}.
\end{equation}
To formally justify this choice, we note that taking fully homogeneous resources yields a solution where the fluctuations (over $i$) of harvest values $h_i$ scale as $\frac{1}{\sqrt N}$. This is necessarily the case when $h$ is determined from an optimization problem of the form~\eqref{eq:optimize}, with the integration measure $\mu\propto\exp(N F(h))$. To probe non-trivial behavior, an externally imposed perturbation of $h_i$ (imposed via $R_i$) must be of the same order. The purpose of this section is to supplement this formal argument with some intuition about what this scaling ansatz encodes.

Intuitively, the form~\eqref{eq:scalingR} may appear restrictive, describing a close-to-homogeneous resource supply. However, in any simulation (or any real ecosystem) $N$ is finite, and the relevant question is the range of values of $\dRSq$ for which our analytical results provide a good approximation. Empirically, the results hold far beyond the range $\dRSq\simeq 1$ assumed by the derivation. We illustrate this by plotting the number of species coexisting at equilibrium for $N=100$ as a function of $\delta R=\sqrt{\dRSq}$, for the extreme (maximally heterogeneous) case of a step-like resource perturbation (with $R_{1,\dots,50}=1+\frac{\delta R}{\sqrt{100}}$ and $R_{51,\dots,100}=1-\frac{\delta R}{\sqrt{100}}$):

\begin{figure}[h!]
\center
  \includegraphics[width=0.4\textwidth]{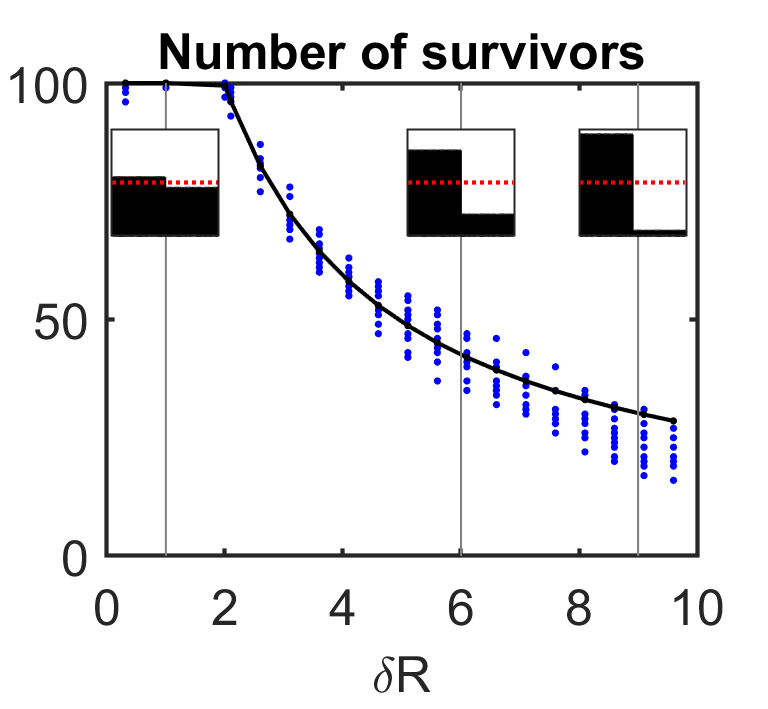}
\caption{Number of survivors for a strongly heterogeneous resource supply. Datapoints show simulation results for $N=100$, with first and last 50 resources supplied at $1\pm\frac{\delta R}{\sqrt N}$. Insets illustrate the supply heterogeneity at three specific values ($\delta R=1$, 6 and 9; dotted line indicates mean supply, i.e. 1). Solid line shows theoretical prediction~\cite{PRL} computed under the ``small-fluctuation'' assumption. The curve provides an excellent approximation up to $\delta R=6$, and remains reasonably accurate even at extreme values of resource supply heterogeneity. For $\delta R<1.98$ the number of survivors hits $N$ (the V/S phase transition~\cite{PRL}).}
\label{fig:S1}
\end{figure}

We see that the theoretical prediction computed under the ``small-fluctuation assumption'' provides an excellent approximation up to $\delta R\approx 6$, and remains reasonably accurate even at $\delta R$ approaching the largest possible value $\sqrt N=10$.

This ``surprising'' agreement can be understood as follows. The most important qualitative implication of the ``small-fluctuation'' scaling adopted here is that no resource is ever depleted to 0. For the resource depletion model introduced in the main text, $h_i(T_i) = \frac{R_i}{T_i}$ remains always positive, and our results extrapolate well. For other models of resource supply, however, a complete depletion of over-exploited resources becomes a possibility. For instance, in his original formulation of the model, MacArthur considered resources renewing at a finite rate. In this case the function $h_i(T_i)$ (the dependence of resource availability on total demand) takes a linear form $h_i(T_i) = a-b T_i$~\cite{PRL}; in particular, it hits zero at a finite demand $T_i$. In this regime, our small-fluctuation theory will no longer be correct, and we refer the reader to Ref.~\cite{Advani} where it is studied.

The large-fluctuation regime is no longer universal, and different resource supply models are no longer equivalent. For our purposes, however, this additional complication appears unnecessary. Indeed, all the effects considered here stem from the feedback of organisms onto their environment. Since the qualitative properties of evolutionary sequences described here are already observed for weak feedback, it is sufficient to consider this simpler scenario. Stronger feedback, capable of fully depleting a certain resource, can be expected to only increase the relevance of our findings.

\subsection{Cost model: approximate neutrality}
In our model, a species with a strategy vector $\sigma^*_i$ is assigned a cost:
\[
\chi^*=\sum_i\sigma^*_i+\epsilon x^*
\]
with a normally distributed $x^*$ and a small $\epsilon$. This cost model corresponds to the assumption of approximate neutrality. Indeed, setting $\epsilon=0$ (at finite $N$) yields a fully degenerate scenario where setting $h_i\equiv 1$ satisfies the resource balance of any species. In other words, for $\epsilon=0$ any combination of species where the total supply matches the total demand ($T_i=R_i$, $N$ equations for $S$ variables) constitutes a dynamical equilibrium. Any non-zero $\epsilon$ breaks this degeneracy, reducing the $S-N$-dimensional space to a single true equilibrium.

Assuming a small $\epsilon$ places us close to the neutral regime. This assumption stems, in part, from the empirical observation of large ecological diversity, which has long motivated neutral or neutral-like models of ecology~\cite{Neutral}. An intuitive argument suggests that evolution might be driving systems to ``emergent neutrality'': organisms that are obviously inferior competitors are eliminated, leaving a self-selected pool of species with approximately similar efficiency. In the context of the model considered here, this argument is detailed in the Supplemental material to Ref.~\cite{PRL}, section S6. For a purely ecological discussion, the nearly-neutral assumption thus appears reasonable~\cite{PRL,eLife}.

When the model is extended to include evolutionary mechanisms, as done here, this empirically motivated shortcut becomes somewhat unsatisfactory. Ultimately, any model adding evolution to a close-to-neutral ecological model should \emph{explain} how it might be driven into and stabilized in this regime. This is very far from obvious (e.g., see Ref.~\cite{Shoresh}), and has long been an active topic of discussion in the literature under the name ``paradox of the plankton''.

Here, we explicitly choose to sidestep this question. For our purposes, we postulate a cost model placing us close to ecological neutrality, and study evolutionary sequences in this regime.

Note that since the solution for the degenerate regime is $h_i\equiv 1$, for small $\epsilon$ we expect $h_i\approx 1$. It is therefore convenient to work with the shifted variables $g_i\equiv 1-h_i$. Note that this definition corrects the unfortunate choice of scaling made in Ref.~\cite{PRL}, which introduced an extra factor of $1/N$ (Ref.~\cite{PRL}, section S7.1), only to later remove it with another change of variables (section S7.6, \emph{ibid}). Except for this modification, the other notations have been kept consistent with Ref.~\cite{PRL}.

\section{Replica calculation of the 2-point joint distribution}\label{app:R}
\subsection{Introduction: recap of the single-copy calculation}
The calculation we are about to undertake is a more complex (two-copy) version of the calculation performed in~\cite{PRL}, continuing from section S7.1 of the supplemental material to that paper. However, to make the narrative self-sufficient, we begin by briefly recapitulating the logic of that computation. We start from the observation that in the MacArthur model, the ecosystem dynamics possess a Lyapunov function $F(\{n_\mu\})$; for the particular resource supply model considered here, $F$ takes the form $F=\sum_i R_i\log h_i(\{n_\mu\})-\sum_\mu \chi_\mu n_\mu$, where the second sum runs over species in the competitor pool. This function is convex and bounded from above~\cite{PRL}. In other words, locating the equilibrium is an $S$-dimensional  optimization problem, where $S=\alpha N$ is the number of species.

With a little algebra (see Ref.~\cite{PRL} for details), the problem of optimizing a complicated function $F(\{n_\mu\})$ over the species abundance space can be converted into optimizing a simpler function $\tilde F(\{h_i\})$ over a complicated region of the resource availability space. The region in question is exactly the unsustainable region $\Omega$ introduced in the main text (Eq.~\eqref{eq:Omega}:
\[
\Omega = \bigcap_{\mu=1}^{\alpha N}\; \{\vec h \;|\; \vec h \cdot \vec\sigma_\mu\le\chi_\mu\}
\]

In order to locate the maximum of $\tilde F$, we investigate the large-$\beta$ limit of the partition function $Z(\beta)$:
\begin{equation}\label{eq:optimize}
\max \tilde F(\{h_i\}) = \lim_{\beta\rightarrow\infty}\left( \frac {\log Z}\beta\right),\text{ where }Z(\beta)=\int_\Omega\! e^{\beta \tilde F(\{h_i\})}\prod_i dh_i
\end{equation}
The problem of locating the equilibrium of an ecosystem is thus converted into the problem of computing a partition function.

Since we expect $g_i\equiv 1-h_i$ to be small, $\tilde F(\{h_i\})$ can be expanded to second order in $g_i$. In the case considered here, we have $\tilde F=-\sum_i R_i\left[g_i+\frac 12 g_i^2\right]$ (for derivation, see Ref.~\cite{PRL}, section S5.3; there are no cross terms $g_i g_j$, because the availability of each resource is set exclusively by the demand for that same resource). In the interest of conciseness, for now we will simply write $\tilde F=\sum_i \tilde F_i(g_i)$.

Recall now the definition of resource surplus of a species, and the cost model $\chi_\mu=\sum_i \sigma_{\mu i}+\epsilon x_\mu$:
\[
\Delta_\mu\equiv \vec h \cdot \vec\sigma_\mu-\chi_\mu \equiv -\sum_i g_i \sigma_{\mu i}-\epsilon x_\mu
\]
Rather than working with a funny-shaped integration region $\Omega$, we introduce resource surplus as an extra integration variable, constrained to be negative: $\Delta_\mu\le0$.

\[
\begin{aligned}
Z
&= \int_\Omega \prod_i dh_i e^{\beta \tilde F}\\
&= \int_0^\infty \prod_i dh_i e^{\beta \tilde F}\prod_{\mu=1}^{\alpha N} \theta
\left(\chi_\mu-\vec h\cdot \vec\sigma_\mu\right)\\
&= \int_{-\infty}^1 \prod_i dg_i e^{\,\beta \tilde F(\{g_i\})}\prod_{\mu=1}^{\alpha N} \int\! d\Delta_\mu\, \theta(-\Delta_\mu)\,
\delta\left(\Delta_\mu+\epsilon x_\mu + \sum_i g_i\sigma_{\mu i}\right)\\
&= \int_{-\infty}^1 \prod_i dg_i e^{\,\beta \tilde F(\{g_i\})}
\prod_{\mu=1}^{\alpha N} \int\! \frac{d\Delta_\mu\, d\hat\Delta_\mu}{2\pi}\,
\theta(-\Delta_\mu)\exp\left[i\sum_{\mu}\hat\Delta_\mu\left( \Delta_\mu+\epsilon x_\mu+\sum_i g_i\sigma_{\mu i} \right)\right]
\end{aligned}
\]
This partition function was computed in Ref.~\cite{PRL}. Our aim now is to generalize that calculation to allow studying the effect of changing $\alpha$.

\subsection{Generalizing the single-copy calculation}
Consider a system of $\alpha_1N$ species. We would like to understand what will happen when the number of species is increased to $\alpha_2N$ with $\alpha_2>\alpha_1$. To do so, we formally consider \emph{two} ecosystems, one with $\alpha_1N$ species, and another with $\alpha_2 N$, the first $\alpha_1 N$ of which are identical to the species in the ecosystem \#1. Denoting $Z_1$ and $Z_2$ the partition functions for each of the two copies, we seek to compute $\langle \log (Z_1 Z_2)\rangle_{x_\mu,\sigma_{\mu i}}$, where the angular brackets denote the averaging over ``disorder'', namely the particular realization of the organisms' costs $x_\mu$ and metabolic strategies $\sigma_{\mu i}$. Using the replica trick:
\begin{equation}\label{eq:trick}
\langle \log Z\rangle  = \lim_{n\rightarrow 0} \frac {\left\langle Z^n\right\rangle-1}n
\end{equation}
we replace this problem by the (much easier) task of computing $\langle Z_1^n Z_2^n\rangle_{x_\mu,\sigma_{\mu i}}$, which we will do for an integer $n$, but then reinterpret $n$ as being a real number one can send to zero. The validity of this procedure, as always, is justified \emph{a posteriori} by the agreement with simulations.

Now that we formally have two coupled ecosystems, we denote the resource availability in each as $h_{ik}\equiv 1-g_{ik}$ and the resource surplus of each species as $\Delta_{\mu k}$, where the index $k$ takes two values, $k=1$ or $2$.

With these notations, the individual partition functions for the two systems are given by:
\begin{multline*}
Z_k = \int \prod_i dg_{ik} e^{\,\beta F_i(\{g_{ik}\})}\\
\prod_{\mu=1}^{\alpha_k N} \int\! \frac{d\Delta_\muk\, d\hat\Delta_\muk}{2\pi}\,
\theta(-\Delta_\muk)\exp\left[i\sum_{\mu}\hat\Delta_\muk\left( \Delta_\muk+\epsilon x_\mu+\sum_i g_\ik\sigma_{\mu i} \right)\right].
\end{multline*}

\subsection{Splitting the partition function into parts: $A_i$, $B_1$ and $B_2$}
Introducing replicas, and considering separately the species shared by the two systems, and those unique to the second ecosystem:
\[
\begin{aligned}
\big\langle Z_1^n &Z_2^n\big\rangle_{x_\mu,\sigma_{\mu i}} =
\int\prod_{i,a,k}\left[d g_\ik^a e^{\beta\sum_{i,a,k} F_i(g_\ik^a)}\right]\\
&\times\prod_{\mu=1}^{\alpha_1 N}\left\{\prod_a\left[\frac{d\Delta_{\mu 1}^a\, d\hat\Delta_{\mu 1}^a}{2\pi}\,\frac{d\Delta_{\mu 2}^a\, d\hat\Delta_{\mu 2}^a}{2\pi}\,
 \theta(-\Delta_{\mu 1}^a)\,\theta(-\Delta_{\mu 2}^a)\right]\,e^{i\sum_a\left(\hat\Delta_{\mu1}^a \Delta_{\mu1}^a+\hat\Delta_{\mu 2}^a \Delta_{\mu 2}^a\right)}\right.\\
&\left.\qquad\qquad\qquad\qquad\times
\left\langle
e^{i\epsilon x_\mu\sum_a(\hat\Delta_{\mu1}^a + \hat\Delta_{\mu2}^a)}
\right\rangle_{x_\mu}
\times
\prod_{i}\left\langle
e^{i \sum_{a}\left(\hat\Delta_{\mu1}^a g_{i1}^a+\hat\Delta_{\mu2}^a g_{i2}^a\right)\sigma_{\mu i}}
\right\rangle_{\sigma_{\mu i}}\right\}\\
&\times\prod_{\mu=1+\alpha_1 N}^{\alpha_2 N}\left\{\prod_a\left[\frac{d\Delta_{\mu 2}^a\, d\hat\Delta_{\mu 2}^a}{2\pi}\,
 \theta(-\Delta_{\mu 2}^a)\right]\,e^{i\sum_a\hat\Delta_{\mu 2}^a \Delta_{\mu 2}^a}\right.\\
&\left.\qquad\qquad\qquad\qquad\times
\left\langle
e^{i\epsilon x_\mu\sum_a\hat\Delta_{\mu2}^a}
\right\rangle_{x_\mu}
\times
\prod_{i}\left\langle
e^{i \sum_{a}\hat\Delta_{\mu2}^a g_{i2}^a\sigma_{\mu i}}
\right\rangle_{\sigma_{\mu i}}\right\}
\end{aligned}
\]
Performing the averaging over disorder is no different than in the single-copy calculation~\cite{PRL}, and we find:
\[
\begin{aligned}
\big\langle Z_1^n&Z_2^n\big\rangle_\mathrm{disorder}=
\int\!\prod_{i,a,k}d g_\ik^a e^{\beta\sum_{i,a,k} F_i(g_\ik^a)}
\prod_a\prod_{\text{all valid $(\mu,k)$}}\frac{d\Delta_\muk^a\, d\hat\Delta_\muk^a}{2\pi}\, \theta(-\Delta_\muk^a)\\
&\times\prod_{\mu=1}^{\alpha_1 N}\exp\left\{i\sum_{a,k}\hat\Delta_\muk^a \Big[\Delta_\muk^a+p\sum_i g_\ik^a\Big]-\frac12 \epsilon^2\Big[\sum_{a,k}\hat\Delta_\muk^a\Big]^2-
\frac {p(1-p)}{2}\sum_{i}\Big[\sum_{a,k}\hat\Delta_\muk^ag_\ik^a\Big]^2\right\}\\
&\times\prod_{1+\alpha_1N}^{\alpha_2 N}\exp\left\{i\sum_{a}\hat\Delta_\mutwo^a \Big[\Delta_\mutwo^a+p\sum_i g_\itwo^a\Big]-\frac12 \epsilon^2\Big[\sum_{a}\hat\Delta_\mutwo^a\Big]^2-
\frac {p(1-p)}{2}\sum_{i}\Big[\sum_{a}\hat\Delta_\mutwo^ag_\itwo^a\Big]^2\right\}
\end{aligned}
\]
Our next step is to decouple indices $i$ and $\mu$ by introducing new variables:
\[
\begin{aligned}
m^a_k&\equiv \sum_i g_{ik}^a\\
q^{ab}_k&\equiv \sum_i g_{ik}^a g_{ik}^b\\
r^{ab}&\equiv \sum_i g_{i1}^a g_{i2}^b
\end{aligned}
\]
Here, $q_1$, $q_2$ as well as $m_1$, $m_2$ are quantities that refer only to a single copy; when it comes to solving saddle-point equations, we will be able to simply substitute the (already established~\cite{PRL}) single-copy values expected for the corresponding $\alpha$. Note that unlike $q_1$, $q_2$, the overlap matrix $r^{ab}$ is \textbf{not} manifestly symmetric in its indices.

Introducing new variables requires inserting a corresponding delta-function:
\begin{multline*}
1=\prod_{a,k}\int\frac{dm^a_k\,d\hat m^a_k}{2\pi}e^{i\hat m^a_k\left(m^a_k-\sum_i g_\ik^a\right)}
\prod_{a,b}\int\frac{dr^{ab}\,d\hat r^{ab}}{2\pi}e^{i\hat r^{ab}\left(r^{ab}-\sum_i g_{i1}^a g_{i2}^b\right)}\\
\times\prod_{a\le b; k}\int\frac{dq^{ab}_k\,d\hat q^{ab}_k}{2\pi}e^{i\hat q^{ab}_k\left(q^{ab}_k-\sum_i g_\ik^a g_\ik^b\right)}
\end{multline*}
We can now factor our integral as follows:
\[
\begin{aligned}
&\left\langle Z_1^nZ_2^n\right\rangle
=\int\prod_{a\le b; k}\frac{dq^{ab}_k\,d\hat q^{ab}_k}{2\pi} \prod_{a,b}\frac{dr^{ab}\,d\hat r^{ab}}{2\pi}
\int\prod_{a,k}\frac{dm^{a}_k\,d\hat m^{a}_k}{2\pi}
\,\exp\; i\left[\sum_{a\le b; k}q^{ab}_k\hat q^{ab}_k+\sum_{a,b}r^{ab}\hat r^{ab}+\sum_{a,k} \hat m^a_k m^a_k\right]\\
&\quad\times\prod_i\left\{
\int_{-\infty}^{1}\prod_{a,k} dg_\ik^a\ \exp\left[\sum_{a,k}\beta F_i(g_\ik^a)-i\sum_{a,k}\hat m^a_k g_\ik^a
-i\sum_{a\le b; k}\hat q^{ab}_kg_\ik^a g_\ik^b -i\sum_{a,b}\hat r^{ab}g_{i1}^a g_{i2}^b \right]\right\}\\
&\quad\times\prod_{\mu=1}^{\alpha_1N}\left\{
\int\prod_{a,k} \frac{d\Delta_\muk^a\,d\hat\Delta_\muk^a}{2\pi}\,\theta(-\Delta_\muk^a) \exp\left[i\sum_{a,k}\hat\Delta_\muk^a(\Delta_\muk^a+pm^a_k)-
\frac12\sum_{a,b,k}\left(p(1-p)q^{ab}_k+\epsilon^2\right)\hat \Delta_\muk^a \hat \Delta_\muk^b\right.\right.\\
&\qquad\qquad\qquad\qquad\qquad\qquad\qquad\qquad\qquad\qquad\qquad\qquad\qquad\qquad
\left.\left.-\sum_{a,b}\left(p(1-p)r^{ab}+\epsilon^2\right)\hat \Delta_{\mu1}^a \hat \Delta_{\mu2}^b
\right]\right\}\\
&\quad\times\prod_{1+\alpha_1N}^{\alpha_2N}
\int\prod_{a} \frac{d\Delta_\mutwo^a\,d\hat\Delta_\mutwo^a}{2\pi}\,\theta(-\Delta_\mutwo^a) \exp\left[i\sum_{a}\hat\Delta_\mutwo^a(\Delta_\mutwo^a+pm^a_2)-
\frac12\sum_{a,b}\left(p(1-p)q^{ab}_2+\epsilon^2\right)\hat \Delta_\mutwo^a \hat \Delta_\mutwo^b\right]\\
\\
&\quad=\int\prod_{a\le b; k}\frac{dq^{ab}_k\,d\hat q^{ab}_k}{2\pi} \prod_{a,b}\frac{dr^{ab}\,d\hat r^{ab}}{2\pi}
\int\prod_{a,k}\frac{dm^{a}_k\,d\hat m^{a}_k}{2\pi}
\,\exp\; i\left[\sum_{a\le b; k}q^{ab}_k\hat q^{ab}_k+\sum_{a,b}r^{ab}\hat r^{ab}+\sum_{a,k} \hat m^a_k m^a_k\right]\\
&\qquad\qquad\qquad\qquad\qquad\qquad\qquad\qquad\qquad\qquad
\times \exp\left[\sum_{i=1}^N \log A_i + \alpha_1N \log B_1+ (\alpha_2-\alpha_1)N\log B_2\right].
\end{aligned}
\]

with $A_i$, $B_1$ and $B_2$ given by:
\[
\begin{aligned}
A_i&=\int_{-\infty}^{1}\prod_{a,k} dg_k^a\ \exp\left[\sum_{a,k}\beta F_i(g_k^a)-i\sum_{a,k}\hat m^a_k g_k^a
-i\sum_{a\le b; k}\hat q^{ab}_k g_k^a g_k^b
-i\sum_{a, b}\hat r^{ab} g_1^a g_2^b\right]\\
B_1&=\int\prod_{a,k} \frac{d\Delta^a_k\,d\hat\Delta^a_k}{2\pi}\theta(-\Delta^a_k) \exp\left[
i\sum_{a,k}\hat\Delta^a_k(\Delta^a_k+pm^a_k)
-\frac12\sum_{a,b,k}\left(p(1-p)q^{ab}_k+\epsilon^2\right)\hat \Delta^a_k \hat \Delta^b_k\right.\\
&\qquad\qquad\qquad\qquad\qquad\qquad\qquad\qquad\qquad\qquad\qquad\qquad
\left.
-\sum_{a,b}\left(p(1-p)r^{ab}+\epsilon^2\right)\hat \Delta^a_1 \hat \Delta^b_2
\right]\\
B_2&=\int\prod_a \frac{d\Delta^a\,d\hat\Delta^a}{2\pi}\theta(-\Delta^a) \exp\left[i\sum_a\hat\Delta^a(\Delta^a+pm^a_2)-
\frac12\sum_{a,b}\left(p(1-p)q^{ab}_2+\epsilon^2\right)\hat \Delta^a \hat \Delta^b\right]
\end{aligned}
\]
At a fully replica-symmetric saddle point, we must have $m^a_k=m^*_k$ (similarly for $\hat m^a_k$). For two-index quantities we distinguish the diagonal and off-diagonal parts, e.g.:
\[
\hat q^{ab}=\left\{\begin{aligned}
&\hat q_D&&\text{ if $a=b$}\\
&\hat q_O&&\text{ if $a\neq b$}
\end{aligned}\right.
\]
This yields:
\begin{multline*}
\log \langle Z_1^nZ_2^n\rangle =
\mathrm{extr}\;
\Big\{\sum_k \Big[in q^D_k\hat q^D_k + i\frac{n(n-1)}{2} q^O_k \hat q^O_k+in \hat m^*_k m^*_k\Big]
+in r^D\hat r^D\\
+ in(n-1) r^O \hat r^O+\sum_i\log A_i + \alpha_1N\log B_1 + (\alpha_2-\alpha_1)N\log B_2\Big\}.
\end{multline*}
Introduce rescaled notations:
\[
\begin{aligned}
i\hat m^*_k&\equiv \beta\hat m_k\\
m_k^*&\equiv m_k\\
q^D_k\approx q^O_k &\equiv q_k\\
r^D\approx r^O &\equiv r\\
q^D_k-q^O_k &\equiv \frac{N}{\beta}x_k\\
r^D-r^O &\equiv \frac{N}{\beta}\rho\\
i\left(\hat q_k^D-\frac 12\hat q_k^O\right)&\equiv\beta a_k\\
\sqrt{-i\hat q^O_k} &\equiv \frac{\beta b_k}{\sqrt N}\\
i\left(\hat r^D-\hat r^O\right) &\equiv \beta \gamma\\
\sqrt{-i\hat r^O} &\equiv \frac {\beta \delta}{\sqrt N}
\end{aligned}
\]
These are similar to the ones used in~\cite{PRL}, except with new variables ($r$, $\rho$, $\gamma$ and $\delta$) added to describe the coupling between our two ecosystems. Note that the somewhat non-orthodox scaling of some of the variables is retained to preserve consistency with Ref.~\cite{PRL}. In these notations, and taking the limit $n\rightarrow 0$:
\begin{multline}\label{eq:extr}
\langle \log Z_1Z_2\rangle =
\lim_{n\rightarrow0}\mathrm{extr}\; \beta
\Big\{\sum_k\left[a_k q_k - \frac12 b_k^2x_k + \hat m_k m_k\right]+\gamma r - \rho\delta^2 \\
+ \frac 1{n\beta}\sum_i\log A_i +\frac {\alpha_1N}{\beta n}\log B_1 +\frac {(\alpha_2-\alpha_1)N}{\beta n}\log B_2\Big\}.
\end{multline}
The calculation of $A_i$ and $B_2$ in this expression is a straightforward generalization of the one performed in~\cite{PRL}, and we will do this first (sections~\ref{sec:A},~\ref{sec:B2}). The final ingredient, namely the expression for $B_1$, will take slightly more effort, and will be the subject of sections~\ref{sec:B1} through~\ref{sec:X}.

\subsection{Computing $\log A_i$}\label{sec:A}
\subsubsection{Symmetric resources}
First, let us assume for simplicity that the supply of all resources is identical $R_i\equiv \bar R=1$. We then have $A_i\equiv A$:
\[
A=\int_{-\infty}^{1}\prod_{a,k} dg_k^a\ \exp\left[\sum_{a,k}\beta F(g_k^a)-i\sum_{a,k}\hat m^*_k g_k^a
-i\sum_{a\le b; k}\hat q^{ab}_k g_k^a g_k^b
-i\sum_{a, b}\hat r^{ab} g_1^a g_2^b\right]
\]
Recall now the expression for the resource supply function $F(g)=-g-\frac 12 g^2$ to find:
\[
A=\int_{-\infty}^{1}\prod_{a,k} dg_k^a\ \exp\left[-\beta \sum_{a,k}g_k^a-\frac{\beta}{2}\sum_{a,k}(g_k^a)^2-i\sum_{a,k}\hat m^*_k g_k^a
-i\sum_{a\le b; k}\hat q^{ab}_k g_k^a g_k^b
-i\sum_{a, b}\hat r^{ab} g_1^a g_2^b\right]
\]
We now write (leaving the subscript $k$ implicit):
\[
\begin{aligned}
\sum_{a\le b}\hat q^{ab}g^a g^b
&=\hat q^D\sum_a(g^a)^2+\frac 12 \hat q^O \sum_{a\neq b}g^ag^b\\
&=\left(\hat q^D-\frac {\hat q^O}2\right) \sum_a(g^a)^2+\frac {\hat q^O}2 \Big(\sum_a g^a\Big)^2\\
\sum_{a, b}\hat r^{ab}g_1^a g_2^b
&=(\hat r^D-\hat r^O)\sum_a g^a_1g^a_2 + \hat r^O\Big(\sum_ag_1^a\Big)\Big(\sum_ag_2^a\Big)\\
&=(\hat r^D-\hat r^O)\sum_a g^a_1g^a_2 + \frac{\hat r^O}2\left[
\Big(\sum_a(g_1^a+g_2^a)\Big)^2-\Big(\sum_ag_1^a\Big)^2-\Big(\sum_ag_2^a\Big)^2
\right]\\
\end{aligned}
\]
Substituting this into the expression for $A$:
\begin{multline*}
A=\int_{-\infty}^{1}\prod_{a,k} dg_k^a\ \exp\left\{\sum_{a,k}
\left[-\beta g_k^a-\frac\beta{2} (g_k^a)^2-i\hat m^*_k g_k^a-i\Big(\hat q_k^D-\frac12\hat q_k^O\Big)(g_k^a)^2\right]\right.\\
\left.-i(\hat r^D-\hat r^O)\sum_a g_1^a g_2^a
-i\frac{\hat r^O}2 \Big(\sum_a(g_1^a+g_2^a)\Big)^2
-i\sum_k\left[ (\hat q^O_k-r^O)\frac{(\sum_a g_k^a)^2}2\right]\right\}
\end{multline*}
In rescaled notations:
\begin{multline*}
A=\int_{-\infty}^{1}\prod_{a,k} dg_k^a\ \exp\left\{\sum_{a,k}
\left[-\beta g_k^a-\frac\beta{2} (g_k^a)^2-\beta \hat m_k g_k^a-\beta a_k(g_k^a)^2\right]
-\beta\gamma \sum_a g_1^a g_2^a\right.\\
\left.+\frac12 \left(\frac{\beta\delta}{\sqrt N}\sum_a(g_1^a+g_2^a)\right)^2
+\sum_k\frac12\left(\frac{\beta\sqrt{b_k^2-\delta^2}}{\sqrt N}\sum_a g_k^a\right)^2\right\}
\end{multline*}
In this expression, the first terms have the desired form $\sum_a(\dots)$ (uncoupled replicas). To deal with the last two, we use Feynman's trick of introducing extra Gaussian variables:
\[\exp\left(\frac 12 (Cx)^2\right)=\int\DD z\, e^{z\,Cx}\]
(the curly $\DD$ denotes the standard Gaussian measure with variance 1). This lets us write:
\begin{multline*}
A=\int\DD w_1 \DD w_2 \DD u \prod_{a=1}^n
\int_{-\infty}^{1}dg_1\, dg_2\ \exp\beta\left\{-\sum_k\left(a_k+\frac 12\right)(g_k)^2-\gamma g_1 g_2\right.\\
-\left.\frac1{\sqrt N}\sum_k\Big[u\delta+w_k\sqrt{b_k^2-\delta^2}+\sqrt N(1+\hat m_k)\Big]g_k\right\}
\end{multline*}
As in Ref.~\cite{PRL}, we shift the variable $\hat m_k$ by setting $\hat m_k\equiv -1+\frac{\delta\hat m_k}{\sqrt N}$ (note that the ``1'' in this substitution is actually the average resource supply $\bar R$, which we set to 1). This scaling ansatz will later be verified by the extremum condition for $\delta\hat m$.
\[
\begin{aligned}
A&=\int\DD w_1\, \DD w_2\, \DD u \prod_{a=1}^n
\int_{-\infty}^{1}dg_1 dg_2\ \exp\beta\left\{-\sum_k\left(a_k+\frac 12\right)(g_k)^2-\gamma g_1 g_2\right.\\
&\qquad\qquad\qquad\qquad\qquad\qquad
-\left.\frac1{\sqrt N}\sum_k\Big[u\delta+w_k\sqrt{b_k^2-\delta^2}+\delta\hat m_k\Big]g_k\right\}\\
&=\int\DD w_1\,\DD w_2\,\DD u\,\left[\int_{-\infty}^1 d\mathbf{g}\,\exp\beta\left(
-\frac 12 \mathbf{g}^T \mathbf{M}\mathbf{g} -\frac1{\sqrt N}\, \mathbf{g}^T \cdot\mathbf{v}\right)\right]^n
\end{aligned}
\]
where $\mathbf{g}\equiv(g_1,g_2)$ and
\[
\mathbf{M}=
\begin{pmatrix}
 2a_1+1 & \gamma \\
 \gamma & 2a_2+1 \\
\end{pmatrix}
\]
\[
\mathbf{v}=
\begin{pmatrix}
 v_1\\
 v_2\\
\end{pmatrix}
\equiv
\begin{pmatrix}
 \delta\hat m_1 + w_1\sqrt{b_1^2-\delta^2}+u\delta \\
 \delta\hat m_2 + w_2\sqrt{b_2^2-\delta^2}+u\delta \\
\end{pmatrix}
\]
Conveniently, for small $n$:
\begin{multline*}
\log\int\mathcal\!Dz\, x^n=\log\left[\int\!\mathcal Dz\, (1+n\log x+\dots)\right]\\
=\log \left[1+n\int\!\mathcal Dz\,\log x+\dots\right]=
n\int\!\mathcal Dz\,\log x+\dots
\end{multline*}
Therefore:
\begin{multline*}
\lim_{n\rightarrow0}\frac {\log A}n =
\int\! \DD w_k \,\DD u\, \log
\int_{-\infty}^1\! d\mathbf{g}\,e^{\beta\big[
-\frac 12 \mathbf{g}^T \mathbf{M}\mathbf{g} +\frac1{\sqrt N}\, \mathbf{g}^T \cdot\mathbf{v}\big]}\\
= \frac{\beta}{N}\int\mathcal Dw_1\,\mathcal Dw_2\,\mathcal Du\,\left(
\frac 12 \mathbf{v}^T \mathbf{M}^{-1}\mathbf{v}\right)
\end{multline*}
This is a Gaussian integral of a quadratic form, and is easily computed:
$$
\lim_{n\rightarrow0}\frac {\log A}n = \frac\beta{2N}
\frac{(2a_1+1)(b_2^2+\delta\hat m_2^2)+(2a_2+1)(b_1^2+\delta\hat m_1^2)
-2\gamma(\delta^2+\delta\hat m_1\delta\hat m_2)}{(2a_1+1)(2a_1+1)-\gamma^2}
$$
As in the single-copy calculation~\cite{PRL}, the extremum conditions with respect to $\delta\hat m_k$ set $\delta \hat m_k = 0$, and the expression for the partition function simplifies to:
$$\begin{aligned}
\langle \log Z_1Z_2\rangle &=
\lim_{n\rightarrow0}\mathrm{extr}\; \beta
\Big\{\sum_k\left[a_k q_k - \frac12 b_k^2x_k -m_k\right]+\gamma r - \rho\delta^2 \\
&+\frac12\frac{(2a_1+1)b_2^2+(2a_2+1)b_1^2
-2\gamma\delta^2}{(2a_1+1)(2a_2+1)-\gamma^2}
+\frac {\alpha_1N}{\beta n}\log B_1 +\frac {(\alpha_2-\alpha_1)N}{\beta n}\log B_2\Big\}.
\end{aligned}
$$
The variables $a_k$, $b_k$, $\gamma$, $\delta$ do not appear in $B_1$ or $B_2$. Consequently, the extremum conditions for these variables can already be computed, and these variables eliminated:
\begin{multline*}
\mathrm{extr}_{a,b,\gamma,\delta}\;
\Big\{
\sum_k\Big[a_k q_k - \frac{b_k^2x_k}2\Big]+\gamma r - \rho\delta^2
+\frac12\frac{(2a_1+1)b_2^2+(2a_2+1)b_1^2
-2\gamma\delta^2}{(2a_1+1)(2a_2+1)-\gamma^2}\Big\}\\
=-\frac{q_1+q_2}2+\frac 12 \frac{q_1 x_2+q_2x_1-2r\rho}{x_1x_2-\rho^2}
\end{multline*}
The partition function is now a function of only $m_k$, $q_k$, $x_k$, $r$, and $\rho$:
\begin{multline*}
\langle \log Z_1Z_2\rangle =
\mathrm{extr}\; \beta
\Big\{\sum_k \left[-\frac{q_k}2-m_k\right]+\frac 12 \frac{q_1 x_2+q_2x_1-2r\rho}{x_1x_2-\rho^2}+\\
\frac {\alpha_1N}{\beta n}\log B_1 +\frac {(\alpha_2-\alpha_1)N}{\beta n}\log B_2
\Big\}
\end{multline*}

\subsubsection{Restoring generality}
In the interest of clarity, the calculation above assumed fully symmetric resources. Restoring full generality renders the intermediate expressions slightly more complicated, but is fairly straightforward, so we summarize it briefly. The exact same steps lead us to the following expression:
$$
\frac{\log A_i}n = \frac{\beta}{N}\int\mathcal Dw_1\,\mathcal Dw_2\,\mathcal Du\,\left(
\frac 12 \mathbf{v}_i^T \mathbf{M}_i^{-1}\mathbf{v}_i\right)\\
$$
where
$$
\mathbf{M}_i=
\begin{pmatrix}
 2a_1+R_i & \gamma \\
 \gamma & 2a_2+R_i \\
\end{pmatrix}
$$
$$
\mathbf{v}_i=
\begin{pmatrix}
 v_1\\
 v_2\\
\end{pmatrix}
\equiv
\begin{pmatrix}
 \delta\hat m_1 + \delta R_i + w_1\sqrt{b_1^2-\delta^2}+u\delta \\
 \delta\hat m_2 + \delta R_i + w_2\sqrt{b_2^2-\delta^2}+u\delta \\
\end{pmatrix}.
$$
As a sanity check, note that setting $R_i\equiv 1$, with $\delta R_i\equiv 0$ yields the expression for the homogeneous case. Taking the Gaussian integral is again straightforward, at which point we recall that what we need is not $A_i$ individually, but the sum $\sum_i \frac{\log A_i}n$. After summation over $i$, the term linear in $\delta\hat m$ cancels to leading order in $N$; the expansion starts with the quadratic term. The extremum condition will therefore set $\delta\hat m_k=0$ as before. Also to leading order in $N$, in the expression for $M_i$ the difference between $R_i$ and $\bar R=1$ is negligible. Integrating away $a_k$, $b_k$, $\gamma$, $\delta$ as before, we obtain our final expression, which now includes two new terms, each proportional to $\dRSq$:
$$
\boxed{
\begin{aligned}
\langle \log Z_1Z_2\rangle =
\mathrm{extr}\; \beta
\Big\{\sum_k &\left[-\frac{q_k}2-m_k+\frac{\dRSq}2x_k\right]+\frac 12 \frac{q_1 x_2+q_2x_1-2r\rho}{x_1x_2-\rho^2}
\\&\qquad+\dRSq\rho+\frac {\alpha_1N}{\beta n}\log B_1 +\frac {(\alpha_2-\alpha_1)N}{\beta n}\log B_2
\Big\}
\end{aligned}}
$$

\subsection{Computing $\log B_2$}\label{sec:B2}
Recall the definition of $B_2$:
$$
B_2=\int\prod_a \frac{d\Delta^a\,d\hat\Delta^a}{2\pi}\theta(-\Delta^a) \exp\left[i\sum_a\hat\Delta^a(\Delta^a+pm^a_2)-
\frac12\sum_{a,b}\left(p(1-p)q^{ab}_2+\epsilon^2\right)\hat \Delta^a \hat \Delta^b\right]
$$
This is a single-copy expression: it pertains to the species that are present only in one of the ecosystems. Not surprisingly, therefore, this $B_2$ is an exact match to an expression already computed in Ref.~\cite{PRL}, where it was called simply $B$ with no subscripts. We can therefore write the result directly:
$$
\lim_{n\rightarrow0}\frac {\log B_2}n=
-\frac {\beta\psi_2^2}{2Nx_2p(1-p)} \int_{\lambda_2}^\infty\!\mathcal Dw\, (w-\lambda_2)^2
=-\frac {\beta\psi_2^2}{2Nx_2\,p(1-p)} I(\lambda_2),
$$
where $\psi_2\equiv\sqrt{p(1-p)q_2+\epsilon^2}$, $\lambda_2\equiv pm_2/\psi_2$ and $I(\lambda)$ is given by:
$$
I(\lambda)\equiv\int_\lambda^\infty e^{-\frac{w^2}2}(w-\lambda)^2\frac{dw}{\sqrt{2\pi}} = -\frac{\lambda}{\sqrt{2\pi}}e^{-\frac{\lambda^2}2}+\frac {1+\lambda^2}2\erfc\left(\frac \lambda{\sqrt2}\right).
$$
\subsection{Computing $\log B_1$ (and introducing $\mathcal X$)}\label{sec:B1}
So far our calculation was a straightforward generalization of Ref.~\cite{PRL}. The only thing that remains is $B_1$, and this is the part that requires some effort:
\begin{multline*}
B_1=\int\prod_{a,k} \frac{d\Delta^a_k\,d\hat\Delta^a_k}{2\pi}\theta(-\Delta^a_k) \exp\left[
i\sum_{a,k}\hat\Delta^a_k(\Delta^a_k+pm^a_k)\right.\\
\left.
-\frac12\sum_{a,b,k}\left(p(1-p)q^{ab}_k+\epsilon^2\right)\hat \Delta^a_k \hat \Delta^b_k
-\sum_{a,b}\left(p(1-p)r^{ab}+\epsilon^2\right)\hat \Delta^a_1 \hat \Delta^b_2
\right]
\end{multline*}

Under replica-symmetric assumptions:
$$
\begin{aligned}
&\sum_{ab}\left[p(1-p)r^{ab}+\epsilon^2\right] \hat\Delta_1^a\hat\Delta_2^b\\
&=p(1-p)(r_D-r_O)\sum_a\hat\Delta_1^a\hat\Delta_2^a+\left[p(1-p)r_O+\epsilon^2\right]
\Big(\sum_a \hat \Delta_1^a\Big)\Big(\sum_b \hat \Delta_2^{b}\Big)\\
&=p(1-p)(r_D-r_O)\sum_a\hat\Delta_1^a\hat\Delta_2^{a}
+\frac{p(1-p)r_O+\epsilon^2}2
\left[
\left(\sum_a \hat \Delta_1^a+\sum_a \hat \Delta_2^a\right)^2 - \Big(\sum_a \hat \Delta_1^a\Big)^2
- \Big(\sum_a \hat \Delta_2^a\Big)^2\right].
\end{aligned}
$$
By analogy with $\psi\equiv\sqrt{p(1-p)q+\epsilon^2}$, introduce a yet another notation:
$$
\sqrt{p(1-p)r+\epsilon^2}\equiv\zeta\qquad\qquad\sqrt{p(1-p)(q_k-r)}\equiv\phi_k.
$$
The exponent in $B_1$ becomes:
$$
\begin{aligned}
\exp&\left\{
-\frac12\frac{Np(1-p)}{\beta}\sum_a\left[x_1(\hat\Delta_1^a)^2+2\rho\hat \Delta_1^a\hat\Delta_2^a +x_2(\hat\Delta_2^a)^2\right]\right.
\\
&+\left.i\sum_{a,k} \hat\Delta_k^a(\Delta_k^a+pm_k^a)-\frac12 \sum_k\phi_k^2 \Big(\sum_a \hat \Delta_k^a\Big)^2-\frac12 \zeta^2 \left(\sum_a \left[\hat \Delta_1^a+\hat \Delta_2^a\right]\right)^2
\right\}
\end{aligned}
$$
Decoupling replicas by introducing auxiliary Gaussian fields:
\begin{align}
B_1&=\int\mathcal Dw_1\,\mathcal Dw_2\,\mathcal Du\int\prod_{a,k} \frac{d\Delta_k^a \,d\hat\Delta_k^a}{2\pi}\theta(-\Delta_k^a)\nonumber\\
&\quad\times\prod_a \exp\Big\{
-\frac{Np(1-p)}{2\beta}\left[x_1(\hat\Delta_1^a)^2+2\rho\hat\Delta_1^a\hat\Delta_2^a +x_2(\hat\Delta_2^a)^2\right]
+i\sum_k\hat\Delta_k^a\left[\Delta_k^a+pm_k^a+w_k\phi_k+u\zeta\right]\Big\}\nonumber\\
&=\int\mathcal Dw_1\,\mathcal Dw_2\,\mathcal Du\left[
\int_{-\infty}^0 \frac{d\DDelta}{2\pi}\,
\int \frac{d\hat\DDelta}{2\pi}\,
\,\exp\left(-\frac12 \frac{Np(1-p)}{\beta} \hat\DDelta^T \mathbf{M}\hat\DDelta +i\hat\DDelta^T\cdot \mathbf{v}\right)\right]^n\nonumber\\
&=\int\mathcal Dw_1\,\mathcal Dw_2\,\mathcal Du\left[\frac 1{\sqrt{\det \mathbf{M}}}
\int_{-\infty}^0 \frac{d\DDelta}{2\pi}\,
\exp\left(-\frac12 \frac{\beta}{Np(1-p)}\mathbf{v}^T \mathbf{M}^{-1}\mathbf{v}\right)\right]^n\label{eq:B1}
\end{align}
where $\DDelta\equiv(\Delta,\Delta')$ (both with and without the ``hats''), and
$$
\mathbf{M}
\equiv
\begin{pmatrix}
 x_1 & \rho \\
 \rho & x_2 \\
\end{pmatrix},
$$
$$
\mathbf{v}
\equiv
\begin{pmatrix}
 \Delta_1+pm_1+w_1\phi_1+u\zeta \\
 \Delta_2+pm_2+w_2\phi_2+u\zeta \\
\end{pmatrix}.
$$
As a result, and neglecting additive terms of lower order:
$$
\begin{aligned}
\lim_{n\rightarrow0}\frac{\log B_1}n &= \int\mathcal Dw_1\,\mathcal Dw_2\,\mathcal Du\, \log
\int_{-\infty}^0 \frac{d\DDelta}{2\pi}\,
\exp\left(-\frac12 \frac{\beta}{Np(1-p)}\mathbf{v}^T \mathbf{M}^{-1}\mathbf{v}\right)\\
&= \frac{\beta}{2Np(1-p)} \int\mathcal Dw_1\,\mathcal Dw_2\,\mathcal Du\, \max_{\Delta_1,\Delta_2\le0}
\left(-\mathbf{v}^T \mathbf{M}^{-1}\mathbf{v}\right)
\end{aligned}
$$

To avoid carrying around the overall multiplication factor, let's rewrite the partition function a little bit, and state this as an \textbf{intermediate result}:
\begin{equation}\label{eq:intermediate}
\boxed{
\begin{aligned}
\langle \log Z_1 Z_2\rangle =
\frac12\mathrm{extr}\; \beta
\Big\{\sum_k &\left[-q_k-2m_k+\dRSq x_k\right]+\frac{q_1 x_2+q_2x_1-2r\rho}{x_1x_2-\rho^2}\\
&+2\dRSq\rho-\frac {\alpha_2-\alpha_1}{p(1-p)x_2}\, \psi_2^2I(\lambda_2)-\frac {\alpha_1}{p(1-p)}\mathcal X
\Big\}
\end{aligned}
}
\end{equation}

The $\mathcal X$ in the last term is given by:
$$
\mathcal X = \int\mathcal Dw_1\,\mathcal Dw_2\,\mathcal Du\, \min_{\Delta_1,\Delta_2\le0}
\left(\mathbf{v}^T \mathbf{M}^{-1}\mathbf{v}\right),
$$
where
$$
\mathbf{M}\equiv
\begin{pmatrix}
 x_1 & \rho \\
 \rho & x_2 \\
\end{pmatrix},\;
\vv\equiv
\begin{pmatrix}
 \vv_1\\
 \vv_2\\
\end{pmatrix}
=
\begin{pmatrix}
 \Delta_1+pm_1+w_1\phi_1+u\zeta \\
 \Delta_2+pm_2+w_2\phi_2+u\zeta \\
\end{pmatrix},
$$
and it would perhaps be helpful to remind the reader of our notations:
$$
\zeta\equiv\sqrt{p(1-p)r+\epsilon^2}\qquad\qquad\phi_k\equiv\sqrt{p(1-p)(q_k-r)}.
$$
The full expression for $\mathcal X$ is rather complicated, but we will now make an important observation that will simplify our life dramatically.
\subsection{Simplifying observation: $\rho=0$}
Recall the original definition of $r^{ab}$:
$$
r^{ab}\equiv \sum_i g_{i1}^a g_{i2}^b
$$
Note, however, that the numbering of the two sets of replicas (for one ecosystem and for the other) is arbitrary, and we can reorder one set without changing the other. It follows that at the saddle point, all the entries in this matrix must be the same (this is the parameter we called $r$). In particular, there can be no difference between the diagonal and the off-diagonal entries, and therefore the variable we called $\rho\equiv\frac\beta N(r^D-r^O)$ must vanish at the saddle point.

This observation means that we only need to calculate $\mathcal X$ up to linear order in $\rho$. Indeed, to determine $r$ all we need is the saddle-point condition for $\rho$, which is computed at $\rho=0$:
$$
\left.\frac{\partial}{\partial\rho}\right|_{\rho=0}\langle\log (Z_1 Z_2)\rangle=0\quad\Rightarrow\quad\text{will determine $r$}.
$$

\subsection{Final ingredient: computing $\mathcal X$ to linear order in $\rho$}\label{sec:X}
Let us rotate the basis of the Gaussian noise variables to introduce:
$$
\begin{aligned}
\xi_1&=pm_1+w_1\phi_1+u\zeta\\
\xi_2&=pm_2+w_2\phi_2+u\zeta
\end{aligned}
$$
(note that these are \emph{not} orthogonal, and therefore these Gaussian variables are no longer independent). We choose the third basis vector to be of unit length, and orthogonal to both $\xi_1$ and $\xi_2$; its explicit expression need not be specified. Since the integrand only involves $\xi_1$ and $\xi_2$, this third Gaussian noise variable can be integrated out, and we are left with:
$$
\mathcal X = \int\tilde{\mathcal D}\vec{\xi}\, \min_{\Delta_1,\Delta_2\le0}
\left[(\DDelta+\vec{\xi})^T \mathbf{M}^{-1}(\DDelta+\vec{\xi})\right],\quad\text{where }\mathbf{M}^{-1} = \begin{pmatrix}\frac1{x_1}&-\frac{\rho}{x_1 x_2}\\-\frac{\rho}{x_1 x_2}&\frac1{x_2}\end{pmatrix}+o(\rho).
$$
Here $\vec\xi\equiv\begin{pmatrix}\xi_1\\ \xi_2\end{pmatrix}$ and the integration measure $\tilde{\DD}\vec\xi$ is that of two correlated Gaussian variables:
$$
\tilde{\DD}\vec\xi=\frac{1}{2\pi\sqrt{\det \mathbf{C}}}\exp\left\{-\frac12 (\vec\xi-p\vec m)^T
\mathbf{C}^{-1}(\vec{\xi}-p\vec m)\right\},
$$
with the correlation matrix:
$$
C=
\begin{pmatrix}
 \phi_1^2+\zeta^2 & \zeta^2 \\
 \zeta^2 & \phi_2^2+\zeta^2 \\
\end{pmatrix}\equiv
\begin{pmatrix}
 \psi_1^2 & \zeta^2 \\
 \zeta^2 & \psi_2^2 \\
\end{pmatrix},\quad\text{where $\psi_k=\sqrt{p(1-p)q_k+\epsilon^2}$ as before.}
$$

Minimizing a quadratic form is not hard; the only complication is the negativity constraint on $\Delta_1$, $\Delta_2$. Globally, the global minimum is always zero, but depending on the values of $\xi_1$, $\xi_2$, the point where it is achieved may lie outside of the allowed quadrant, in which case we will have to content ourselves with the smallest value at the quadrant boundary (where either $\Delta_1$ or $\Delta_2$ is zero). We must therefore consider several cases, and the expression for the integrand will be different. Specifically, we find that the integration plane splits into 4 regions we label A, B, C and D:
$$
\begin{array}{l|l}
\text{Region B: }
\left\{\begin{aligned}
&\xi_1<0\\
&\xi_2-\frac{\rho}{x_1}\xi_1>0
\end{aligned}\right.\quad
&
\text{Region A: }
\left\{\begin{aligned}
&\xi_1>0\\&\xi_2>0\\
\end{aligned}\right.\quad
\\\\
\text{Integrand: }\frac{\xi_1^2}{x_1}
&
\text{Integrand: }0\\
\\
\hline
\\
\text{Region C: }
\left\{\begin{aligned}
&\xi_1-\frac{\rho}{x_2}\xi_2<0\\
&\xi_2-\frac{\rho}{x_1}\xi_1<0
\end{aligned}\right.\quad
&
\text{Region D: }
\left\{\begin{aligned}
&\xi_1-\frac{\rho}{x_2}\xi_2>0\\
&\xi_2<0\\
\end{aligned}\right.\quad
\\\\
\text{Integrand: }\frac{\xi_1^2}{x_1}+\frac{\xi_2^2}{x_2}-2\rho\frac{\xi_1\xi_2}{x_1x_2}
&
\text{Integrand: }\frac{\xi_2^2}{x_2}\\
\end{array}
$$

To get some reassurance we are on the right track, note that setting $\rho=0$ decouples the integrals over $\xi_1$ and $\xi_2$, and we immediately find:
\begin{equation}\label{eq:zeroOrder}
\left.\mathcal X\right|_{\rho=0} =
\int_{\{\xi_1<0\}}\tilde{\mathcal D}\vec{\xi}\,\frac{\xi_1^2}{x_1}+
\int_{\{\xi_2<0\}}\tilde{\mathcal D}\vec{\xi}\,\frac{\xi_2^2}{x_2}=
\frac{\psi_1^2}{x_1}\,I\left(\frac{pm_1}{\psi_1}\right)+\frac{\psi_2^2}{x_2}\,I\left(\frac{pm_2}{\psi_2}\right).
\end{equation}
This is precisely the expected result. Indeed, substituting this into our expression~\eqref{eq:intermediate} yields
$$
\langle \log Z_1 Z_2\rangle =
\sum_k\mathrm{extr}\; \beta
\Big\{-\frac{q_k}2-m_k+\frac{\dRSq}2 x_k+\frac{q_k}{x_k}-\frac{\alpha_k\psi_k^2}{p(1-p)x_k}I\left(\frac{pm_k}{\psi_k}\right)\Big\},
$$
which is reassuring, as we of course expect $\langle \log Z_1 Z_2\rangle =\langle \log Z_1 \rangle + \langle \log Z_1 \rangle$.

To determine $r$, we now need to calculate the first-order correction to \eqref{eq:zeroOrder} and find the relevant saddle-point equation $\left.\frac{\partial}{\partial\rho}\right|_{\rho=0}(\dots)=0$. A non-zero $\rho$ brings two differences: first, the integration region is not quite the same, and second, there is an extra term in the integrand. Let us write this as follows:

$$
\begin{aligned}
\mathcal X &=
\int_{C+B}\tilde{\mathcal D}\vec{\xi}\,\frac{\xi_1^2}{x_1}+
\int_{C+D}\tilde{\mathcal D}\vec{\xi}\,\frac{\xi_2^2}{x_2}-
2\rho\int_{C}\tilde{\mathcal D}\vec{\xi}\,\frac{\xi_1\xi_2}{x_1x_2}+o(\rho)\\\\
&=\int_{C+B-\{\xi_1<0\}+\{\xi_1<0\}}\tilde{\mathcal D}\vec{\xi}\,\frac{\xi_1^2}{x_1}
+\int_{C+D-\{\xi_2<0\}+\{\xi_2<0\}}\tilde{\mathcal D}\vec{\xi}\,\frac{\xi_2^2}{x_2}-2\rho\int_{C}\tilde{\mathcal D}\vec{\xi}\,\frac{\xi_1\xi_2}{x_1x_2}+o(\rho)\\\\
&=\frac{\psi_1^2}{x_1}\,I\left(\frac{pm_1}{\psi_1}\right)+\frac{\psi_2^2}{x_2}\,I\left(\frac{pm_2}{\psi_2}\right)\\
&\qquad\qquad+\int_{C+B-\{\xi_1<0\}}\tilde{\mathcal D}\vec{\xi}\,\frac{\xi_1^2}{x_1}+
\int_{C+D-\{\xi_2<0\}}\tilde{\mathcal D}\vec{\xi}\,\frac{\xi_2^2}{x_2}-
2\rho\int_{C}\tilde{\mathcal D}\vec{\xi}\,\frac{\xi_1\xi_2}{x_1x_2}+o(\rho)
\end{aligned}
$$
(Here we add and subtract integration regions in the obvious sense of adding/subtracting the integrals taken over them.) The terms in the last line all vanish at $\rho=0$: the first two, because the integration region vanishes, and the final one, because it is explicitly multiplied by $\rho$. Conveniently, therefore, only the very last term contributes to the derivative $\left.\frac{\partial}{\partial\rho}\right|_{\rho=0}$, and we find:
$$
\left.\frac{\partial}{\partial\rho}\right|_{\rho=0} \mathcal X =
\left.\frac{\partial}{\partial\rho}\right|_{\rho=0} \left[
-2\rho\int_C \tilde\DD\vec\xi \, \frac{\xi_1\xi_2}{x_1x_2}
\right]=- \frac{2}{x_1x_2}\int_{\xi_1,\xi_2<0} \tilde\DD\vec\xi \, \xi_1\xi_2
$$
Recalling Eq.~\eqref{eq:intermediate}, we write the saddle-point equation:
$$
\begin{aligned}
0&=\left.\frac{\partial}{\partial\rho}\right|_{\rho=0}\langle\log Z_1Z_2\rangle = \left.\frac{\partial}{\partial\rho}\right|_{\rho=0}
\left[\frac12\frac{q_1 x_2+q_2x_1-2r\rho}{x_1x_2-\rho^2}+\dRSq\rho-\frac {\alpha_1}{2p(1-p)}\mathcal X\right]\\
&=-\frac{r}{x_1x_2}+\dRSq+\frac {\alpha_1}{p(1-p)x_1x_2}\int_{\xi_1,\xi_2<0} \tilde\DD\vec\xi \, \xi_1\xi_2
\end{aligned}
$$
Changing the sign of the integration variable, we find the equation for $r$:
$$
r = \dRSq x_1x_2+\frac{\alpha}{p(1-p)}\int_{0}^\infty
\frac{d\xi_1\,d\xi_2}{2\pi\sqrt{\det C}}\exp\left\{-\frac12 (\vec\xi+p\vec m)^T [C(r)]^{-1}(\vec\xi+p\vec m)\right\}
$$
where
$$
C(r)=
\begin{pmatrix}
 \psi_1^2 & \zeta^2 \\
 \zeta^2 & \psi_2^2 \\
\end{pmatrix}\equiv
\begin{pmatrix}
 p(1-p)q_1+\epsilon^2 & p(1-p)r+\epsilon^2 \\
 p(1-p)r+\epsilon^2 & p(1-p)q_2+\epsilon^2
\end{pmatrix},
$$
and $x_k=1-\frac{\alpha_k}2\erfc\left(\frac{pm_k}{\psi_k\sqrt 2}\right)$ at the saddle point (see Eq.~(S17) in Ref.~\cite{PRL}). The version quoted in the main text assumed $\dRSq=0$ for simplicity.

\subsection{The joint distribution $p(\Delta_1, \Delta_2)$}
Having computed $r$, we will now demonstrate how this parameter is related to the shape of the joint distribution $p(\Delta_1, \Delta_2)$ (for species that are common to both ecosystems). For this, we go back to the expression for $B_1$ given in Eq.~\eqref{eq:B1}. Instead of directly sending $n\rightarrow0$, we separate out one of the replicas, leaving $n-1$ remaining ones, and only then send $n$ to zero (for a more detailed explanation of this trick, see Ref.~\cite{PRL}, section S9). We find:
$$
\begin{aligned}
p(\Delta_1,\Delta_2)&=\int\mathcal Dw_1\,\mathcal Dw_2\,\mathcal Du\,\left[
\frac{\exp\left(-\frac\beta 2\mathbf{v}^T \mathbf{M}^{-1}\mathbf{v}\right)}{\int_{\Delta_{1,2}<0} d\Delta_1 d\Delta_2\exp\left(-\frac\beta2 \mathbf{v}^T \mathbf{M}^{-1}\mathbf{v}\right)}
\right]\\
&\equiv \int\mathcal Dw_1\,\mathcal Dw_2\,\mathcal Du\,\,p(\Delta_1,\Delta_2|w_1,w_2,u).
\end{aligned}
$$
where
$$
\mathbf{M}\equiv Np(1-p)
\begin{pmatrix}
 x_1 & \rho \\
 \rho & x_2 \\
\end{pmatrix},\;
\vv\equiv
\begin{pmatrix}
 \vv_1\\
 \vv_2\\
\end{pmatrix}
=
\begin{pmatrix}
 \Delta_1+pm_1+w_1\phi_1+u\zeta \\
 \Delta_2+pm_2+w_2\phi_2+u\zeta \\
\end{pmatrix},
$$
and $\phi_k=\sqrt{p(1-p)(q_k-r)}$. The key observation is that the conditional distribution $p(\Delta_1,\Delta_2|w_1,w_2,u)$ is manifestly a properly normalized probability distribution on the negative quadrant $\Delta_{1,2}<0$. For any $\{w_1, w_2,u\}$, the limit of $p(\Delta_1,\Delta_2|w_1,w_2,u)$ as $\beta\rightarrow\infty$ is therefore easy to compute: it is always a (properly normalized) delta-function, concentrated where the quadratic form in the exponent reaches its minimum, which could be either within the quadrant, or at its boundary.

Let us determine the portion of $p(\Delta_1,\Delta_2)$ for purely negative $\Delta_1$, $\Delta_2$. As $\beta\rightarrow\infty$, the conditional distribution $p(\Delta_1,\Delta_2|w_1,w_2,u)$ concentrates into a 2-dimensional delta-function:
$$
p(\Delta_1,\Delta_2|w_1,w_2,u)\stackrel{\beta\rightarrow\infty}{\longrightarrow}
\delta(\Delta_1+pm_1+w_1\phi_1+u\zeta)\,\delta(\Delta_2+pm_2+w_2\phi_2+u\zeta)
$$
Therefore, for strictly negative $\Delta_1,\Delta_2$:
$$
p(\Delta_1,\Delta_2)=\int\mathcal Dw_1\,\mathcal Dw_2\,\mathcal Du\, \delta(\Delta_1+pm_1+w_1\phi_1+u\zeta)\,\delta(\Delta_2+pm_2+w_2\phi_2+u\zeta).
$$
Introduce new noise variables $\xi_1=w_1\phi_1+u\zeta$ and $\xi_2=w_2\phi_2+u\zeta$. The noise direction orthogonal to both can be integrated away and gives 1. What remains is a 2-dimensional integral:
$$
p(\Delta_1,\Delta_2)=\int\mathcal D\xi_1\,\mathcal D\xi_2\,\, \delta(\Delta_1+pm_1+\xi_1)\,\delta(\Delta_2+pm_2+\xi_2),
$$
where the measure is that of two correlated Gaussian variables:
$$
\DD\xi_1\,\DD\xi_2=\frac{1}{2\pi\sqrt{\det C(r)}}\exp\left\{-\frac12 \vec\xi^t\cdot \left[C(r)\right]^{-1}\cdot \vec\xi\right\}
\quad
\text{where}
\quad
C(r)=
\begin{pmatrix}
 \psi_1^2 & \zeta^2 \\
 \zeta^2 & \psi_2^2.
\end{pmatrix}
$$
($C$ depends on $r$ through $\zeta$.) The result, obviously, is a double Gaussian with the exact same correlation structure, for variables $\Delta_{1,2}+pm_{1,2}$:
$$
p(\Delta_1,\Delta_2)=\frac{1}{2\pi\sqrt{\det C}}\exp\left\{-\frac12 (\mathbf{\Delta}+p\mathbf{m})^t\cdot \left[C(r)\right]^{-1}\cdot (\mathbf{\Delta}+p\mathbf{m})\right\}
$$
This expression is valid only for strictly negative $\Delta_{1,2}$. We find that, as claimed in the main text, the variable $r$ indeed controls the degree of correlation between $\Delta_1$ and $\Delta_2$ (the resource surplus of the same species between the two ecosystems).

Following the same procedure, one can verify that the probability weight at the quadrant boundary is precisely the weight of the same double-Gaussian expression integrated over the forbidden (positive) range of $\Delta$, also as claimed. For instance:
$$
p(\Delta_1<0,\Delta_2=0)=\int_0^\infty d\Delta_2\,\,\frac{1}{2\pi\sqrt{\det C}}\exp\left\{-\frac12 (\mathbf{\Delta}+p\mathbf{m})^t\cdot \left[C(r)\right]^{-1}\cdot (\mathbf{\Delta}+p\mathbf{m})\right\}.
$$
The simplest way to prove this is to observe that the distribution we computed must have the marginal computed in Ref.~\cite{PRL}, and shown in Fig.~2(a).

\section{Correlation of cost and resource surplus (Fig. 3(c))}\label{app:Corr}
In this appendix, we compute the correlation of a species' cost with the resource surplus it experiences. Our goal is to prove the claim made in Fig. 3(c), namely that
for $\alpha$ below a critical value (called the V phase in Ref.~\cite{PRL}), this correlation vanishes. In other words, we will show that in the V phase, the cost of a species has no effect on its survival (the survivors are those species whose resource surplus is zero rather than negative), illustrating the non-intuitive nature of the ``selection pressure'' in this phase of our model.

The computation in this section is a single-system calculation, i.e. we are working with a single $\alpha$, and only one set of species. Our starting point is section S7.2 of Ref.~\cite{PRL} (here and below, all section numbers refer to the supplemental material of Ref.~\cite{PRL}). In order to compute $\langle\sum_\mu x_\mu\Delta_\mu\rangle$, we need to add to the partition function a generating term $\log Z \rightarrow \log Z+\eta \sum_\mu x_\mu\Delta_\mu,$ so that we have
$$
\sum_\mu x_\mu\Delta_\mu = \left.\frac{\partial}{\partial\eta}\right|_{\eta=0} \log Z.
$$
The extra term modifies the result of averaging over the ``disorder'' $x_\mu$, namely the expression which was labeled as (1) in section S7.2. The modified expression is:
$$
(1)\equiv
\prod_\mu\left\langle
e^{i\epsilon \sum_{a}\hat\Delta_\mu^a x_\mu+\eta\sum_a\Delta_\mu^a x_\mu}
\right\rangle_{x_\mu}=\exp\left[-\frac 12 \sum_\mu\Big(\sum_a\epsilon\hat\Delta_\mu^a-i\eta\Delta_\mu^a\Big)^2\right].
$$
This causes a corresponding modification in the expression for $B$ (section S7.3):
$$
\begin{aligned}
B=\int\prod_a \frac{d\Delta^a\,d\hat\Delta^a}{2\pi}\prod_a\theta(-\Delta^a) \exp\Big\{&i\sum_a\hat\Delta^a(\Delta^a+pm^a)\\
&-\frac12 p(1-p)\sum_{a,b}q^{ab}\hat \Delta^a \hat \Delta^b
-\frac12\Big[\sum_a\big(\epsilon\hat\Delta^a-i\eta\Delta^a\big)\Big]^2\Big\}
\end{aligned}
$$
Introducing a replica-symmetric ansatz and decoupling replicas with the Feynman trick, we write:
\begin{equation}\label{eq:smallB}
B=\int\DD w\,\DD w'\, \Big[b(w, w', \eta)\Big]^n
\end{equation}
where
$$
\begin{aligned}
b(w,w', \eta)=
\int\frac{d\Delta\,d\hat\Delta}{2\pi}\theta(-\Delta) \exp\Big\{&i\hat\Delta(\Delta+pm)
-\frac12 p(1-p)(q_D-q_O))\hat \Delta^2 \\
&+i\Big[w\sqrt{p(1-p)q_O}\hat\Delta+w'\big(\epsilon\hat\Delta-i\eta\Delta\big)\Big]\Big\}
\end{aligned}
$$
Grouping together the terms with $\hat\Delta$ and taking that integral:
$$
\begin{aligned}
b(w,w', \eta)=
\int_{-\infty}^0 d\Delta\,\,e^{w'\eta\Delta}&\int\frac{d\hat\Delta}{2\pi} \exp\Big\{
-\frac12 p(1-p)(q_D-q_O))\hat \Delta^2\\ &\qquad\qquad\qquad+ i\hat\Delta\Big(\Delta+pm+w\sqrt{p(1-p)q_O}+w'\epsilon\Big)\Big\}\\
=\int_{-\infty}^0 d\Delta\,\,e^{w'\eta\Delta}\,\,&\frac{1}{\sqrt{2\pi}}
\frac 1{\sqrt{p(1-p)(q_D-q_O)}}
\exp\left[
-\frac{\Big(\Delta+pm+w\sqrt{p(1-p)q_O}+w'\epsilon\Big)^2}{2p(1-p)(q_D-q_O)}
\right]
\end{aligned}
$$
In the limit $n\rightarrow 0$, from Eq.~\eqref{eq:smallB} it follows:
$$
\log B=\log\int\DD w\,\DD w'\, b^n = n\int\DD w\,\DD w' \log b(w,w',\eta)+\dots,
$$
and therefore
$$\left.\frac{\partial}{\partial\eta}\right|_{\eta=0} \log B \\
= n \int\DD w\,\DD w'\,\, \frac{\int_{-\infty}^0 d\Delta\,
\frac{w'\Delta}{\sqrt{2\pi p(1-p)(q_D-q_O)}}
\exp\Big\{
-\frac{\big[\Delta+pm+w\sqrt{p(1-p)q_O}+w'\epsilon\big]^2}{2p(1-p)(q_D-q_O)}
\Big\}
}{b(w,w',0)}.
$$
Omitting the constants that cancel out, recalling that $q_D-q_O =\frac{N}{\beta}x$, and rescaling the integration variable by a factor $\sqrt{p(1-p)x}$, we can write:
$$
\langle x\Delta\rangle = \int\DD w\,\DD w'\,\frac{b_1(w,w')}{b_0(w,w')},
$$
where the numerator and the denominator are given by:
$$
\begin{aligned}
b_1(w,w')&=\int_{-\infty}^0\frac{d\Delta}{\sqrt{2\pi}}\exp\left\{
-\frac12\frac{\beta}{N}\left(\Delta+\Delta_0\right)^2\right\}\times w'\Delta\sqrt{p(1-p)x}\\
b_0(w,w')&=\int_{-\infty}^0\frac{d\Delta}{\sqrt{2\pi}}\exp\left\{
-\frac12\frac{\beta}{N}\left(\Delta+\Delta_0\right)^2\right\}.
\end{aligned}
$$
Here $\Delta_0\equiv\frac{pm+w\sqrt{p(1-p)q}+\epsilon w'}{\sqrt{p(1-p)x}}$. If $\Delta_0$ is positive, then as $\beta\rightarrow\infty$, the exponent tends to a delta function (centered at $-\Delta_0$), and the ratio tends to $-w'\Delta_0\sqrt{p(1-p)x}$. If, however, $\Delta_0$ is negative, then the exponent starts concentrating at $\Delta=0$, where the numerator vanishes. As a result:
$$
\langle x\Delta\rangle = \int\DD w\,\DD w'\,\theta\left(\frac{pm+w\sqrt{p(1-p)q}+\epsilon w'}{\sqrt{p(1-p)x}}\right)\times(-w')(pm+w\sqrt{p(1-p)q}+\epsilon w').
$$
Denoting $F(u)\equiv u\,\theta(u)$ and integrating by parts (note that $w'\DD w'$ is a full derivative):
$$
\begin{aligned}
\langle x\Delta\rangle &= -\int\DD w\,\DD w'\,w'\,F\left(pm+w\sqrt{p(1-p)q}+\epsilon w'\right)\\
&= -\int\DD w\,\DD w'\,\frac{\partial}{\partial w'}\,F\left(pm+w\sqrt{p(1-p)q}+\epsilon w'\right)\\
&= -\epsilon\int\DD w\,\DD w'\,\theta\left(pm+w\sqrt{p(1-p)q}+\epsilon w'\right).
\end{aligned}
$$
Rotating the $(w,w')$ basis to integrate away the one of the noise directions, we are left with a 1-d integral in which we recognize the function $E(\lambda)\equiv \frac 12 \erfc\left(\frac\lambda{\sqrt 2}\right)$, concluding with a very simple formula:
$$
\langle x\Delta\rangle = -\epsilon E\left[\frac{-pm}{\sqrt{p(1-p)q+\epsilon^2}}\right]\equiv-\epsilon E\left[\frac{-pm}{\psi}\right]
$$
The parameter combination $\lambda\equiv pm/\psi$ plays an important role throughout our model; for instance, the number of survivors is $\alpha E(\lambda)$~\cite{PRL}. Here we see that it also controls the correlation of species' cost and resource surplus (and thus their survival as well). To get the correlation coefficient as plotted in Fig.~3(c), we should normalize this result by the standard deviations. The standard deviation of $x$ is 1. As for $\Delta$, its distribution is shown in Fig.~3(a); the standard deviation of the Gaussian part is $\psi$. Neglecting the effect of the delta-shaped tail (whose total weight is at most $1/\alpha$, and whose effect on the variance quickly becomes negligible as $\alpha$ increases), we find:
$$
\mathrm{corr}(x,\Delta)\equiv\frac{\langle x\Delta\rangle}{\sqrt{\langle x^2\rangle}\sqrt{\langle \Delta^2\rangle}} \simeq
-\frac{\epsilon E(-\lambda)}{\psi}
$$
Consider the behavior of this expression as $\epsilon\rightarrow0$; this is the limit where V/S becomes a true phase transition. As established in Ref.~\cite{PRL}, for $\alpha$ below a critical value, $\psi$ remains of order 1, and we find that the correlation vanishes in the V phase, as promised. For $\alpha$ above the critical value, $\psi$ was shown to go to zero linearly with $\epsilon$, so that their ratio is of order 1 and the correlation becomes non-trivial. For small, but finite $\epsilon$ the behavior retains these qualitative features and agrees with simulations, as shown in Fig. 3(c).

\end{document}